\documentclass[showkeys,aps,amssymb,prd]{revtex4}
\usepackage{bm,graphicx}

\newcommand\beq{\begin{equation}}
\newcommand\beqa{\begin{eqnarray}}
\newcommand\beqan{\begin{eqnarray*}}
\newcommand\eeq{\end{equation}}
\newcommand\eeqa{\end{eqnarray}} 
\newcommand\eeqan{\end{eqnarray*}}

\newcommand\gravr{{\sf m}_\bullet}
\newcommand\bhpm{{\cal M}_\bullet}
\newcommand\bha{{\sf a}}
\newcommand\bhpJ{{\cal J}_\bullet}

\newcommand\cL{{\cal L}}

\newcommand\hcL{\hat{\cal L}}
\newcommand\hcQ{\hat{\cal Q}}

\newcommand\vthE{\vartheta_E}
\newcommand\vep{\varepsilon}
\newcommand\vphi{\varphi}
\newcommand\vth{\vartheta}
\newcommand\dvphi{\delta\varphi}
\newcommand\dchi{\delta\chi}
\newcommand\vthbh{\vartheta_\bullet}
\newcommand\ath{{\hat{\alpha}_{\text{hor}}}}
\newcommand\atv{{\hat{\alpha}_{\text{vert}}}}

\newcommand\sns{{\tt s}}
\newcommand\snq{{\tt q}}

\newcommand\ssc{{\sf c}}
\newcommand\ssd{{\sf d}}
\newcommand\ssf{{\tt F}}
\newcommand\ssg{{\tt G}}

\newcommand\cb{{\cal B}}

\newcommand\cd{{\tt d}}
\newcommand\order[2]{\mathcal{O}\left({#1}\right)^{#2}}
\newcommand\ahat{\hat{a}}

\newcommand\kpolar{\zeta}
\newcommand\kazym{\phi}

\newcommand\reffig[1]{Fig.~\ref{fig:#1}}

\newcommand\refsec[1]{Section~\ref{sec:#1}}

\newcommand\refapp[1]{Appendix~\ref{app:#1}}

\begin{document}

\title{Lensing by Kerr Black Holes. II: Analytical Study of Quasi-Equatorial Lensing Observables}

 \author{Amir B.\ Aazami}
\affiliation{Department of Mathematics,\\ Duke University,\\
  Science Drive, Durham, NC 27708-0320;\\
  {\tt aazami@math.duke.edu}}

\author{Charles R.\ Keeton}
\affiliation{Department of Physics \& Astronomy, Rutgers University,
  136 Frelinghuysen Road, Piscataway, NJ 08854-8019;\\
  {\tt keeton@physics.rutgers.edu}}

\author{A.\ O.\ Petters}
\affiliation{Departments of Mathematics and Physics, Duke University,\\
  Science Drive, Durham, NC 27708-0320;\\
  {\tt petters@math.duke.edu}}
  
\begin{abstract}
In this second paper, we develop an analytical theory of quasi-equatorial lensing by Kerr black holes.  In this setting we solve perturbatively our general lens equation with displacement given in Paper I, going beyond weak-deflection Kerr lensing to third order in our expansion parameter $\vep$, which is the ratio of
the angular gravitational radius to the angular Einstein radius.  We obtain
new formulas and results for the bending angle, image positions, image magnifications, total unsigned
magnification, and centroid, all to third order in $\vep$ and including the displacement.  New results on the time delay between images are also given to second order in $\vep$, again including displacement.  For all lensing observables we show that the displacement begins to appear only at second order in $\vep$.  When there is no spin, we obtain new results on the lensing observables for Schwarzschild lensing with displacement.

\end{abstract}
\keywords{black holes, gravitational lensing, lens equation}
\maketitle


\section{Introduction}
\label{sec:intro}

In Paper I \cite{AKP} we derived a general lens equation and magnification formula governing gravitational lensing by Kerr black holes.  Our equation took into account the displacement that arises when the light ray's tangent lines at the source and observer do not intersect on the lens plane.

In this second paper we study the new lens equation of Paper I.  We shall restrict our attention to quasi-equatorial lensing by a Kerr black hole, and  address lensing observables in this regime.  Our Paper II begins by obtaining the full light bending angle
with ``horizontal'' and ``vertical'' components
for an equatorial observer and light rays that are quasi-equatorial.  Next, we develop analytically quasi-equatorial Kerr lensing
beyond weak-deflection Kerr lensing 
to third order in $\vep$, which is the ratio of
the angular gravitational radius to the angular Einstein radius.
Specifically, we solve our lens equation perturbatively to obtain
formulas for the lensing observables: image position, image magnification, total unsigned
magnification, centroid, and time delay.  It is shown that the displacement
begins to affect the lensing observables
only at second order in $\vep$, and so can safely be ignored for
studies of first-order corrections to weak-deflection quasi-equatorial
Kerr lensing.  Finally, the findings in the paper also yield new results on the lensing observables in Schwarzschild lensing with displacement.

The outline of the paper is as follows.  In Section~\ref{sec:results} we carefully list the assumptions used throughout the paper.  In Section~\ref{sec:bend-angle} we determine the ``horizontal" component of the bending angle; the details of the computation itself appear in Appendix~\ref{app:Kerr-bend-angle}.  Section~\ref{sec:lensing} contains the bulk of our results: we derive formulas for image position, magnification, total magnification, centroid, and time delay, all expressed perturbatively
to second or third order
in the expansion parameter $\vep$.  Due to the lengthy forms of the third-order results, we list them in Appendix~\ref{app:third-order}.  Also, as a simple application of our results, we plot the image correction terms as a function of the angular source position and compare with the Schwarzschild case.  Appendix~\ref{sec:app:vertical} contains new results about the vertical bending angle in quasi-equatorial Kerr lensing.

\section{Definitions and Assumptions}
\label{sec:results}

\begin{figure}[t]
\begin{center}
\includegraphics[scale=.62]{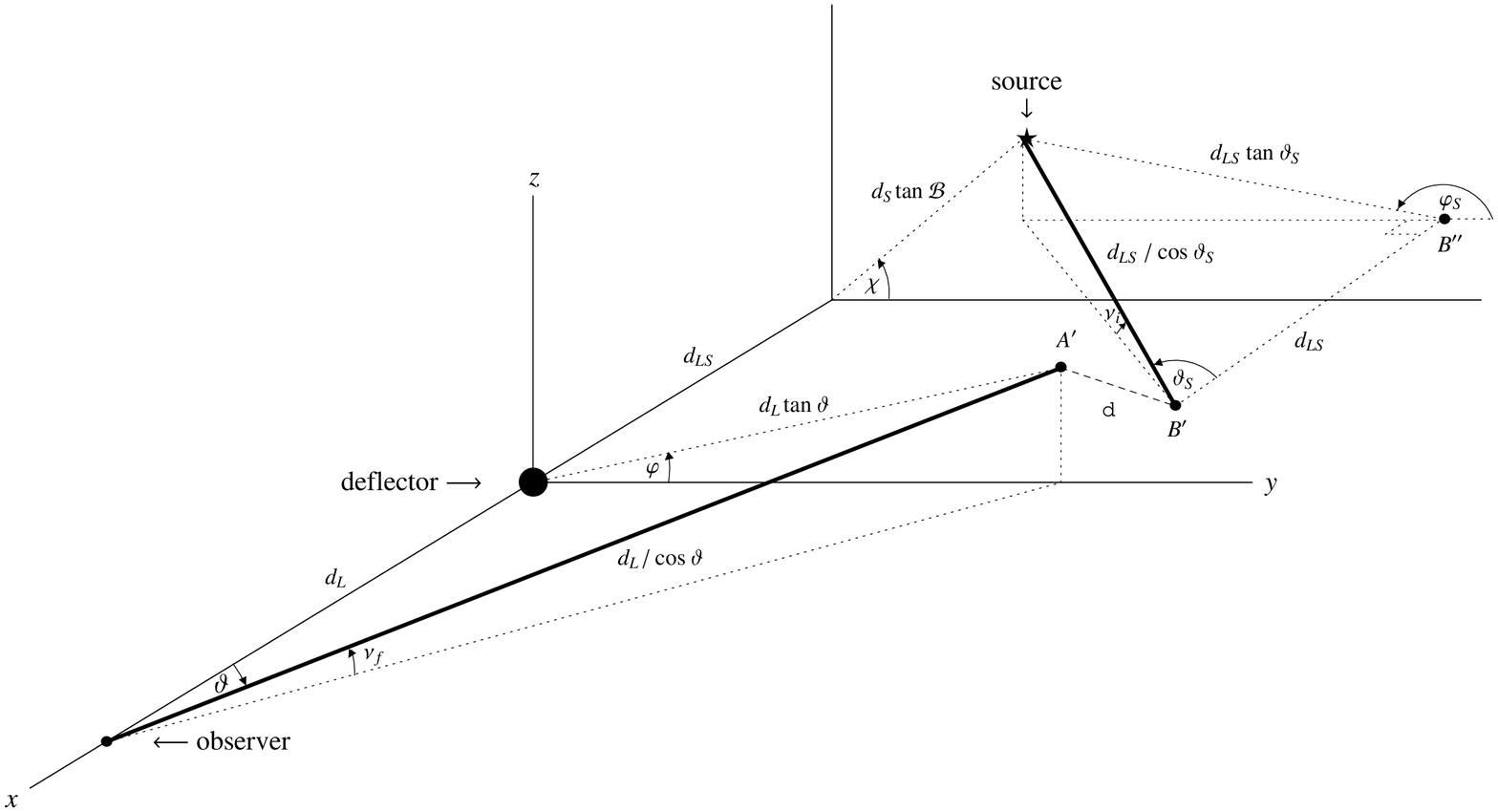}
\end{center}
\caption{
Geometry for lensing with displacement by a Kerr black hole.  Taken from Fig.~3 of Paper I \cite{AKP}.}
\label{fig:vangle}
\end{figure}

We define Cartesian coordinates $(x,y,z)$ centered on the
Kerr black hole and oriented such that the observer lies on the
positive $x$-axis.  As in Paper I, we
introduce spherical polar angles centered on the observer and
defined with respect to the optical axis (the $x$-axis) and the lens and light source planes.  The vector to the image position is described by the
angle $\vth$ it makes with the optical axis and by an azimuthal
angle $\vphi$ in the lens plane.  Recall that $\vth$ is strictly positive and within the interval $(0,\pi/2)$, while $0 \leq \vphi < 2\pi$.  Similarly, the vector to the source position is
described by the angle $\cb$ it makes with the optical axis and by
an azimuthal angle $\chi$ in the light source plane; $\cb$ is also strictly nonnegative and within the interval $[0,\pi/2)$, while $0 \leq \chi < 2\pi$ (see Fig.~1 of Paper I).  A diagram of the lensing geometry is shown in Fig.~\ref{fig:vangle}.  Note that $-\pi/2 < \vth_S < \pi/2$ and $0 \leq \vphi_S < 2\pi$.  We adopt the same sign convention for $\vth_S$ as in Paper I: if $\vth_S$ is measured {\it toward} the optical axis, then it will be positive; otherwise it is negative (e.g., the $\vth_S$ shown in \reffig{vangle} is positive).

We now state the assumptions under which we will work throughout this paper:

\begin{enumerate}
\item The Kerr black hole and the light source are not at cosmological distances, so that $d_S = d_L + d_{LS}$, where $d_S$ and $d_L$ are the perpendicular distances from the observer to the source and lens planes, respectively, and $d_{LS}$ is the perpendicular distance from the lens plane to the source plane;

\item Both the source and observer are in the asymptotically flat region of the Kerr spacetime, and the observer lies in the equatorial plane of the Kerr black hole.  This last condition implies that the coordinates $(x,y,z)$ coincide with the Boyer-Lindquist coordinates $(X,Y,Z)$ centered on the black hole (see \reffig{BHCoords} below);

\item The source is not required to be incrementally close to the optical axis and can be either on the equatorial plane or slightly off it, so that $\chi = \chi_0 + \dchi$, where $\chi_0 = 0$ or $\pi$.  Similarly, the lift of the light ray off the equatorial plane is small, so that $\vphi = \vphi_0 + \dvphi$ and $\vphi_S = \vphi_0 + \pi + \dvphi_S$, where $\vphi_0 = 0$ (retrograde motion) or $\pi$ (prograde motion), and where $\dvphi$ and $\dvphi_S$ are small and considered
only to linear order.  We henceforth refer to this as the {\it quasi-equatorial regime}.
\end{enumerate}

In Paper I we derived the following general lens equation governing lensing by a thin deflector, for source and observer in the asymptotically flat region:
\beqa
d_S \tan\cb \, \cos\chi & = & d_L \tan\vth \, \cos\vphi
  \ + \ d_{LS} \tan \vth_S \, \cos \vphi_S
  \ + \ \cd_y\,,
  \label{eq:lens1h} \\
d_S \tan\cb \, \sin\chi & = & d_L  \tan\vth \, \sin\vphi
  \ + \ d_{LS} \tan \vth_S \, \sin \vphi_S
  \ + \ \cd_z\,.
  \label{eq:lens1v}
\eeqa
Here the displacements are shown explicitly; note that $\left(\cd_y^2+\cd_z^2\right)^{1/2} = \cd$ in \reffig{vangle}.

Specializing to the case of an equatorial observer in the Kerr metric, we also derived in Paper I the following lens equation with the displacements implicitly included: 
\beqa
  d_S \tan\cb \cos\chi &=& d_{LS} \tan\vth_S \cos\vphi_S
    \ + \ d_L\ \frac{\sin\vth \cos\vphi}{\cos\vth_S}\ ,
    \label{eq:le-h-Kerr} \\
  d_S \tan\cb \sin\chi &=& d_{LS} \tan\vth_S \sin\vphi_S
    \ + \ \frac{d_L \sin\vth}{1 - \sin^2\vth_S \sin^2\vphi_S} \times
    \label{eq:le-v-Kerr} \\
  &&\qquad \left[ \cos\vphi \sin\vth_S \tan\vth_S \sin\vphi_S \cos\vphi_S
    + \left( \sin^2\vphi - \sin^2\vth_S \sin^2\vphi_S \right)^{1/2} \right] .
    \nonumber
\eeqa
This is valid for all light rays, whether they loop
around the black hole or not, as long as they lie outside the
region of photon capture.  No small-angle approximation is required.  Comparing our two lens equations allowed us to extract the displacement terms:
\beqa
  \cd_y &=& d_L \sin\vth \cos\vphi \left( \frac{1}{\cos\vth_S}
    - \frac{1}{\cos\vth} \right) , \label{eq:dispy}\\
  \cd_z &=& - d_L \tan\vth \sin\vphi \ + \ 
    \frac{d_L \sin\vth}{1 - \sin^2\vth_S \sin^2\vphi_S} \times \label{eq:dispz}\nonumber\\
  &&\qquad \left[ \cos\vphi \sin\vth_S \tan\vth_S \sin\vphi_S \cos\vphi_S
    + \left( \sin^2\vphi - \sin^2\vth_S \sin^2\vphi_S \right)^{1/2} \right].
\eeqa

\section{Quasi-Equatorial Kerr Light Bending}
\label{sec:bend-angle}

With that as background, we begin this paper by calculating the component of the bending angle in the equatorial plane, which is the $xy$-plane in Fig.~\ref{fig:vangle}; we call this the ``horizontal" component.  Due to the technical nature of the calculations,
we quote the key results here and refer to \refapp{Kerr-bend-angle}
for the detailed treatment.  Note from \reffig{vangle} that 
according to the way the angles $\vth$ and $\vth_S$ are defined, they may lift off the $xy$-plane or a plane parallel to it.  Let us define $\hat{\vth}$ and $\hat{\vth}_S$ to be their projections onto the $xy$-plane, respectively.  Without loss of generality, we choose the same sign conventions for $\hat{\vth}$ and $\hat{\vth}_S$ as we chose for $\vth$ and $\vth_S$, in which case we can unambiguously write the ``horizontal" component of the bending angle as
\beqa
\label{ath1}
\ath = \hat{\vth} + \hat{\vth}_S\ .
\eeqa
Note that the positivity of $\hat{\vth}$ and the fact that the bending angle is nonnegative forces the condition
$$
\hat{\vth}_S \geq -\hat{\vth}\ .
$$
(Indeed, with our signs conventions the condition $\hat{\vth}_S < -\hat{\vth}$ would be equivalent to repulsion of the light ray.)
Writing $\hat{\vth}$ and $\hat{\vth}_S$ in terms of the angles $\vth, \vphi, \vth_S, \vphi_S$, we have
\beqa
\hat{\vth} &=& \tan^{-1}(\tan\vth\cos\vphi)\ ,\label{eq:bavh1}\\
\hat{\vth}_S &=& \tan^{-1}(\tan\vth_S\cos(\pi-\vphi_S))\ .\label{eq:bavh2}
\eeqa
As stated in assumption (3) above, in the quasi-equatorial regime we have
\beq
\label{eq:pert-angles}
  \vphi = \vphi_0 + \dvphi\,, \qquad
  \vphi_S = \vphi_0 + \pi + \dvphi_S\,,
\eeq
where $\vphi_0$ is either 0 (retrograde motion) or $\pi$ (prograde motion), while
$\dvphi$ and $\dvphi_S$ are small and considered
only to linear order.  In this regime eqns.~(\ref{eq:bavh1}) and (\ref{eq:bavh2}) simplify to
\beqa
\hat{\vth} &\approx& \pm\vth\ ,\nonumber\\
\hat{\vth}_S &\approx& \pm\vth_S\nonumber\ .
\eeqa
Since $\hat{\vth}$ and $\hat{\vth}_S$ have the same signs as $\vth$ and $\vth_S$, respectively, we discard the negative solutions, so that eqn.~(\ref{ath1}) reduces to
\beqa
\ath \approx \vth + \vth_S\ ,\quad \vth_S \geq -\vth\,.\label{bavh}
\eeqa 
Thus in the quasi-equatorial regime we may use the full angles $\vth$ and $\vth_S$ in place of their respective projections onto the $xy$-plane.
With that said, we show in 
eqn.~(\ref{eq:app:bangle-b1-kerr}) of Appendix~\ref{app:Kerr-bend-angle} that the ``horizontal" bending
of light has the following invariant series expansion:
\beqa \label{eq:bangle-b2-kerr}
  \ath(b) &=& 
          A_1 \left(\frac{\gravr}{b}\right)
    \ + \ A_2 \left(\frac{\gravr}{b}\right)^2
    \ + \ A_3 \left(\frac{\gravr}{b}\right)^3
    \ + \ A_4 \left(\frac{\gravr}{b}\right)^4
    \ + \ \order{\frac{\gravr}{b}}{5} ,
\eeqa
where
\beqa
  A_1 &=& 4\ ,\\
  A_2 &=& \frac{15\pi}{4} - 4\,\sns\,\ahat\ ,\\
  A_3 &=& \frac{128}{3} - 10\,\pi\,\sns\,\ahat + 4\,\ahat^2\ ,\\
  A_4 &=& \frac{3465\pi}{64} -192\,\sns\,\ahat + \frac{285\pi\,\ahat^2}{16} - 4\,\sns\,\ahat^3\ .\label{eq:Ai}
\eeqa
The variable $\sns$ equals $\pm 1$ depending on whether the light ray undergoes prograde $(+1)$ or retrograde $(-1)$ motion (see eqn.~(\ref{sns}) in Appendix~\ref{app:Kerr-bend-angle} below).  Note that eqns.~(\ref{eq:bangle-b2-kerr})--(\ref{eq:Ai}) are consistent with the bending angle obtained by Iyer \& Hansen \cite{Iyer1} by a different means---note also that their bending angle is consistent with the exact bending angle \cite{Iyer1,Iyer2}.  We remind the reader of our conventions in Paper I.  The parameter $\gravr$ is the gravitational radius and $\bha$ 
is the angular momentum per unit mass,
\beqa \label{eq:agravr}
  \gravr = \frac{G \bhpm}{c^2}\ , \qquad
  \bha = \frac{\bhpJ}{c \bhpm}\ ,
\eeqa
where $\bhpm$ is the mass of the black hole and $\bhpJ$ its spin angular momentum (see, e.g., \cite[pp.~322-324]{wald}).  Note that both $\gravr$ and $\bha$ have dimensions of length.  The quantity $\ahat$ is a dimensionless spin parameter:
\beqa
\label{eq:ahat}
  \ahat = \frac{\bha}{\gravr}\ .\nonumber
\eeqa
Unless stated to the contrary, the black hole's spin is
subcritical; i.e., $\ahat^2 < 1$.  Finally, $b = d_L \sin\vth$ is the impact parameter (see eqn.~(A8) in Paper I), which is a constant of the motion.

When there is no spin, the coefficients reduce to
$A_1 = 4$, $A_2 = 15\pi/4$, $A_3 = 128/3$, and $A_4 = 3465\pi/64$ and recover the
Schwarzschild bending angle in \cite{KP1}.  Also,
eqn.~(\ref{eq:bangle-b2-kerr}) shows that in the weak-deflection
limit (at first order in $\gravr/b$) the Kerr bending angle
agrees with the Schwarzschild bending angle.  The spin enters
only in higher-order correction terms.  The sign is such that
the spin makes the bending angle {\em larger} for light rays
that follow retrograde motion ($\sns = -1$).  This makes sense
intuitively because retrograde rays spend more time in the
presence of the black hole's gravitational pull.


\section{Observable Properties of Lensed Images}
\label{sec:lensing}

In this section we derive asymptotic formulas for image position, image
magnification, total unsigned magnification, centroid, and time delay
for quasi-equatorial Kerr lensing with displacement.

\subsection{Quasi-Equatorial Lens Equation}
\label{subsec:quasi}

We begin with our general lens equation (\ref{eq:lens1h})--(\ref{eq:lens1v}) and insert a 
bookkeeping parameter $\xi$ to monitor the displacement in
either $\cd_y$ or $\cd_z$:
\beqa
d_S \tan\cb \, \cos\chi & = & d_L \tan\vth \, \cos\vphi
  \ + \ d_{LS} \tan \vth_S \, \cos \vphi_S
  \ + \ \xi\,\cd_y\,,
  \label{eq:lens1h} \\
d_S \tan\cb \, \sin\chi & = & d_L  \tan\vth \, \sin\vphi
  \ + \ d_{LS} \tan \vth_S \, \sin \vphi_S
  \ + \ \xi\,\cd_z\,.
  \label{eq:lens1v}
\eeqa
(The displacements $\cd_y$ and $\cd_z$ are given by eqns.~(\ref{eq:dispy}) and (\ref{eq:dispz}).)
We can take $\xi=1$ to include the displacements
properly, or choose $\xi=0$ if we wish to ignore the
displacements (in order to connect with work in \cite{KP1,KP2}).

Beginning with eqn.~(\ref{eq:lens1h}), we substitute eqn.~(\ref{eq:dispy}) in place of $\cd_y$ and Taylor expand in the small angles $\dvphi$, $\dvphi_S$, and $\dchi$, to obtain
\beq
  \snq \tan\cb = (1-D) \tan\vth \ - \ D \tan\vth_S
    \ + \ \xi (1-D) \sin\vth \left( \frac{1}{\cos\vth_S}
      - \frac{1}{\cos\vth} \right) 
  \ + \ \order{2}{} ,
  \label{eq:lens2h}
\eeq
where $\snq = \cos(\chi_0-\vphi_0)$, $D = d_{LS}/d_S$, and
$\order{2}{}$ indicates terms that are second order in
$\dvphi$, $\dvphi_S$, and/or $\dchi$.  (Below, we incorporate
the sign $\snq$ into the tangent so that the left-hand side is
written as $\tan(\snq\cb)$ and we think of $\snq\cb$ as the
signed source position.)  This is the ``horizontal'' component
of the lens equation.   Bear in mind that $\xi$ identifies terms
associated with the displacement.  Including the displacement by setting $\xi=1$ in eqn.~(\ref{eq:lens2h}) yields
\beqa 
\label{eq:LensEqn2a}
  \snq\,\tan\cb = (1-D)\frac{\sin\vth}{\cos\vth_S} - D \tan\vth_S
    \ + \ \order{2}{}.\nonumber
\eeqa
Thus, to lowest order
in out-of-plane motion we recover the same lens equation as
in the Schwarzschild case (see eqn.~(19) of Paper I).

We use $\vth_S = \ath - \vth$, taking $\ath$ from
eqn.~(\ref{eq:bangle-b2-kerr}), and introduce scaled angular
variables:
\beq \label{eq:newvar}
  \beta = \frac{\snq\cb}{\vthE}\ , \quad
  \theta = \frac{\vth}{\vthE}\ , \quad
  \vthbh = \tan^{-1}\left(\frac{\gravr}{d_L}\right)\ ,\quad
  \vep = \frac{\vthbh}{\vthE} = \frac{\vthE}{4\,D}\ .
\eeq
Here the natural angular scale is given by the angular Einstein
ring radius:
\beqa
\label{eq:einrad}
  \vthE = \sqrt{\frac{4 G \bhpm d_{LS}}{c^2 d_L d_S}} = \sqrt{\frac{4 \gravr D}{d_L }}\ .
\eeqa
Note that we have defined the scaled source position $\beta$
to be a signed quantity, with a sign that indicates whether
the source is on the same or opposite side of the lens as the
image.  In eqn.~(\ref{eq:bangle-b2-kerr}) we wrote the bending angle
$\ath$ as a series expansion in $\gravr/b$.  For
analyzing the observable image positions, $\vep$ is the more
natural expansion parameter.  To convert $\ath$ into a series expansion in $\vep$, note that according to eqns.~(\ref{eq:newvar}) and (\ref{eq:einrad}) and the fact that $b = d_L\sin\vth$, we have
\beqa
\label{eq:mb}
\frac{\gravr}{b} = \frac{4D\,\vep^2}{\sin(4D\,\vep\,\theta)} = \frac{1}{\theta}\,\vep + \frac{8D^2\,\theta}{3}\,\vep^3 + \frac{224D^4\,\theta^3}{45}\,\vep^5 + \order{\vep}{7}\ .
\eeqa

As in \cite{KP1,KP2,KP3}, we postulate that the solution of the ``horizontal" lens equation (\ref{eq:lens2h})
can be written as a series expansion of the form
\beq \label{eq:tseries}
  \theta = \theta_0\, +\, \theta_1\,\vep\, +\, \theta_2\,\vep^2\, +\, \theta_3\,\vep^3\, +\,
    \order{\vep}{4}.
\eeq
Converting now to our scaled angular variables (\ref{eq:newvar})--(\ref{eq:tseries}), our quasi-equatorial ``horizontal" lens equation (\ref{eq:lens2h}) takes the form
\beqa
  \beta &=& \left[ \theta_0 - \frac{1}{\theta_0} \right]
    \ + \ \frac{1}{\theta_0^2} \left[\sns\,\ahat - \frac{15\pi}{16}
      + (1+\theta_0^2)\,\theta_1 \right]\,\vep
  \label{eq:lens3h}\\
  && + \ \frac{1}{24\,\theta_0^3} \Biggl[12\,\sns\,\ahat\,(5\pi - 4 \theta_1) - 24\,\ahat^2
      - 384 + 3\,\theta_1\,(15\pi-8\theta_1)
      + 24\,\theta_0\,\theta_2\,(1+\theta_0^2)
      \nonumber\\
  &&\qquad
      +\ 8\,\theta_0\left(48\,D\,\theta_0 + 8D^2\,\theta_0^2 (-2 \beta^3 - 7 \theta_0
      + 2 \theta_0^3)\right)
      + 192\theta_0^2\,(1-D) (1 - 2\,D\,\theta_0^2)\,\xi \Biggr]\,\vep^2
      \nonumber\\
      &&  + \ \frac{1}{768\,\theta_0^4} \Biggl[128\,\sns\,\ahat\left\{384 + 9 \theta_1 (-5 \pi + 2 \theta_1) + 
   4 \theta_0 \left(-3 \theta_2 + 
      8 \theta_0 (2 D (-3 + 2 D \theta_0^2)\right.\right.\nonumber\\ 
        && \qquad+\ \left.\left.\!3 (1 - D) (-1 + D \theta_0^2)\,\xi)\right)\right\} - 768\,\ahat^2\,(2\pi - 3\theta_1) + 768\,\sns\,\ahat^3 \nonumber\\
        &&\qquad+\ 15 \pi \left\{-3 (487 + 48 \theta_1^2) + 
    32\,\theta_0 \left(3 \theta_2 + 
       8 \theta_0 (2 D (3 - 2 D \theta_0^2)\right.\right.\nonumber\\ 
       &&\qquad+\left.\left.\ \!3 (-1 + D) (-1 + D \theta_0^2)\,\xi)\right)\right\} + 
 256 \left\{3 \theta_1^3 + 3 \theta_0^2 (1 + \theta_0^2) \theta_3\right.\nonumber\\
 &&\qquad+\left.\ 
    2 \theta_1 \left(72 + \theta_0 (-3 \theta_2 + 
          4 \theta_0 (D (-6 - 7 D \theta_0^2 + 6 D \theta_0^4) + 
             3 (-1 + D) (1 + 2 D \theta_0^2)\,\xi))\right)\right\}\Biggr]\,\vep^3\nonumber\\
  && + \ \order{\vep}{4} .  \nonumber
\eeqa
Note that displacement terms (indicated by $\xi$) only begin to appear
at second order.  Also, since we are simultaneously expanding $\tan\snq\cb = \tan(4\,\beta\,D\,\vep)$, note the occurrence of $\beta^3$ in the $\vep^2$ term.   


Now we turn to the ``vertical'' component of the lens equation, namely,
eqn.~(\ref{eq:lens1v}).  Substituting eqn.~(\ref{eq:dispz}) in place of $\cd_z$ and Taylor expanding in the small
angles $\dvphi$, $\dvphi_S$, and $\dchi$, we obtain
\beqa
  (\dchi)\,(\snq\tan\cb) &=& (\dvphi)\,(1-\xi) (1-D) \tan\vth
    \ - \ (\dvphi_S)\,D \tan\vth_S
  \label{eq:lens2v}\\
  && + \ \xi (1-D) \sin\vth \left\{ (\dvphi_S \sin\vth_S \tan\vth_S
    + \left[ (\dvphi)^2 - (\dvphi_S)^2 \sin^2\vth_S \right]^{1/2} \right\} 
    \nonumber\\
  && + \ \order{2}{} .
    \nonumber
\eeqa
Next, we use eqn.~(\ref{eq:app:W1}) in Appendix~\ref{app:Kerr-bend-angle} to write $(\dvphi_S)\,\sin\vth_S = (\dvphi)\,W(\vth)$:
\beqa
\label{squareroot}
 (\dchi)\,(\snq\tan\cb) &=& \dvphi\,(1-\xi) (1-D) \tan\vth + \dvphi\,D\,W(\vth) (\cos\vth_S)^{-1}\\
    &+& \xi (1-D) \sin\vth \{ \dvphi\,W(\vth) \tan\vth_S + \dvphi\,[1-W(\vth)^2]^{1/2} \}\ .\nonumber
\eeqa
(In eqn.~(\ref{squareroot2}) of Appendix~\ref{app:Kerr-bend-angle} we show that $1-W(\vth)^2 > 0$, so this equation is never complex-valued.)  Finally, we convert to our scaled angular variables (\ref{eq:newvar})--(\ref{eq:tseries}) and expand in $\vep$, obtaining
\beqa
\label{eq:lens3v}
&&\hspace{-.5in}\left(\beta + \frac{16}{3}D^2\,\beta^3\,\vep^2 + \order{\vep}{4}\right)\dchi = \dvphi\,\left\{\left[\theta_0 - \frac{1}{\theta_0}\right] + \frac{1}{\theta_0^2}\left[2\,\sns\,\ahat -\frac{15\pi}{16} + (1+ \theta_0^2)\,\theta_1\right]\vep \right.\\
&&+~\frac{1}{24\,\theta_0^3}\Biggl[\sns\,\ahat\,(90 \pi - 96\theta_1) + 2 \ahat^2 + 45 \pi \theta_1 
   - 24 (16 + \theta_1^2) + 
   8 \theta_0 \left(8 D^2 \theta_0^3 (-7 + 2 \theta_0^2 + 6 \xi)\right.\nonumber\\ 
   &&\quad+~ 
      \left.\!3 (\theta_2 + \theta_0^2 \theta_2 + 8 \theta_0 \xi) - 
      24 D \theta_0 (-2 + \xi + 
         2 \theta_0^2 \xi)\right)\Biggr]\vep^2\nonumber\\
       &&+~\frac{1}{768\,\theta_0^4}\Biggl[\ahat\,{\tt V_1} + 72\,\ahat^2\,(-95\pi+96\,\theta_1) + 3072\,\sns\,\ahat^3 + {\tt V_2}\Biggr]\vep^3 + \order{\vep}{4}\Biggr\}\ ,\nonumber
\eeqa
where
\beqa
{\tt V_1} &=& 64\,\sns\left[960 + 9 \theta_1 (-15 \pi + 8 \theta_1) + 
 16 \theta_0 \left(-3 \theta_2 + 
    4 \theta_0 (D (-6 + 5 D \theta_0^2) - 
       3 (-1 + D) (-1 + D \theta_0^2) \xi)\right)\right]\ ,\nonumber\\
{\tt V_2} &=& 15\pi\left[-3 (487 + 48 \theta_1^2) + 
 32 \theta_0 \left(3 \theta_2 + 
    8 \theta_0 (2 D (3 - 2 D \theta_0^2) + 
       3 (-1 + D) (-1 + D \theta_0^2) \xi)\right)\right]\nonumber\\
       &&~+256\left[3 \theta_1^3 + 3 \theta_0^2 (1 + \theta_0^2) \theta_3 + 
 2 \theta_1 \left(72 + \theta_0 (-3 \theta_2 + 
       4 \theta_0 (D (-6 - 7 D \theta_0^2 + 6 D \theta_0^4)\right.\right.\nonumber\\
       &&~+\left.\left. 
          3 (-1 + D) (1 + 2 D \theta_0^2) \xi))\right)\right]\ .\nonumber
\eeqa
This is the ``vertical" component of the lens equation.  We will use it to obtain a relation between the small angles $\dchi$ and $\dvphi$.  To that end, we divide eqn.~(\ref{eq:lens3v}) by eqn.~(\ref{eq:lens3h}) to eliminate $\beta$: 
\beqa
  \dchi &=& \dvphi\ \Biggl\{
    1 \ + \ \frac{\sns\,\ahat}{\theta_0(\theta_0^2-1)}\,\vep\label{eq:lens4v}\\
  &&\quad + \ \frac{\ahat}{16 \theta_0^2 (\theta_0^2-1)^2} \Bigl[
      \sns\,(-5\pi + 4 \theta_0^2 (5\pi-12\theta_1)
      + 16\theta_1) + 16 \ahat (1-2\theta_0^2) \Bigr] \vep^2\nonumber\\
      &&\quad +\ \frac{\ahat}{768 \theta_0^3 (\theta_0^2-1)^3} \Bigl[-24\,\ahat\left(\pi(101-262\,\theta_0^2 + 221\,\theta_0^4) - 64(1-3\,\theta_0^2 + 4\,\theta_0^4)\,\theta_1\right)\nonumber\\
      &&\quad +\ 768\,\sns\,\ahat^2\,(1-3\theta_0^2 + 3\,\theta_0^4) + \sns\,{\tt K}\Bigr] \vep^3 + \order{\vep}{4} \Biggr\}\ ,\nonumber
\eeqa
where
\beqa
{\tt K} &=& 225 \pi^2 (-1 + 4 \theta_0^2) - 
  480 \pi (1 - 3 \theta_0^2 + 8 \theta_0^4) \theta_1 + 
  768 \theta_1^2\nonumber\\
  &&~+ 256\left[\theta_0 \left(-3 \theta_2 + \theta_0 (-8 D^2 \
\theta_0^2 (-1 + \theta_0^2) (-5 + 6 \xi) + 
           24 D (-1 + \theta_0^2) (-2 + \xi + 
              2 \theta_0^2 \xi)\right.\right.\nonumber\\
              &&~-\left.\left. 
           3 (16 + 3 \theta_1^2 - 
              8 \xi + \theta_0 (-4 \theta_2 + \theta_0 (-16 - 
                    6 \theta_1^2 + 3 \theta_0 \theta_2 + 
                    8 \xi))))\right)\right]\ .\nonumber
\eeqa
Observe that in general $\dchi \neq \dvphi$ in the regime of quasi-equatorial lensing.  {\em Thus when $\ahat \neq 0$, the light ray's trajectory cannot lie in a plane other than the equatorial plane (in which case $\dvphi = \dchi = 0$).}  Notice that displacement terms (indicated by $\xi$) appear only at third order in $\vep$.

\subsection{Image Positions}

We now solve our ``horizontal" lens equation (\ref{eq:lens3h}) term
by term to find $\theta_0, \theta_1, \theta_2$, and $\theta_3$.  The
zeroth-order term is the familiar weak-deflection lens equation
for the Schwarzschild metric,
\beq \label{eq:lens0}
  \beta = \theta_0 - \frac{1}{\theta_0}\ ,
\eeq
which yields the weak-deflection image position
\beq \label{eq:theta0}
  \theta_0 = \frac{1}{2} \left( \sqrt{\beta^2+4} + \beta \right) .
\eeq
We neglect the negative solution because we have explicitly
specified that angles describing image positions are positive.
For a source with $\beta > 0$, the negative-parity image is obtained by plugging $-\beta$ in eqn.~(\ref{eq:theta0}); note that eqn.~(\ref{eq:theta0}) will still be positive.  (Note also that we are solving for quasi-equatorial images; there may be additional images in the general case.)

Requiring that the first-order term in eqn.~(\ref{eq:lens3h})
vanishes yields
\beq \label{eq:theta1}
  \theta_1 = \frac{15\pi - 16\,\sns\,\ahat}{16 (\theta_0^2+1)}\ .
\eeq
Likewise with the vanishing of the second-order term,
\beqa \label{eq:theta2}
  \theta_2 &=& \frac{1}{24\,\theta_0\,(\theta_0^2+1)} \Biggl[
    64 \left(6 - D (2 D + 6 (1-D) \theta_0^2 - D \theta_0^4)\right)
    + 24 \ahat^2
    - 12\,\sns\,\ahat (5\pi - 4 \theta_1)
    \nonumber\\
  &&\qquad
    - 3 \theta_1 (15\pi-8\theta_1)
    - 192\theta_0^2 (1-D) (1 - 2 D \theta_0^2)\,\xi \Biggr] ,
  \label{eq:theta2}
\eeqa
where we have used eqn.~(\ref{eq:lens0}) to substitute for $\beta$
in terms of $\theta_0$.  Note that the displacement only affects
$\theta_2$, not $\theta_0$ and $\theta_1$.   The third-order image correction term $\theta_3$ is given in Appendix~\ref{app:third-order}.

\begin{figure}[t]
\begin{center}
\includegraphics[scale=.613]{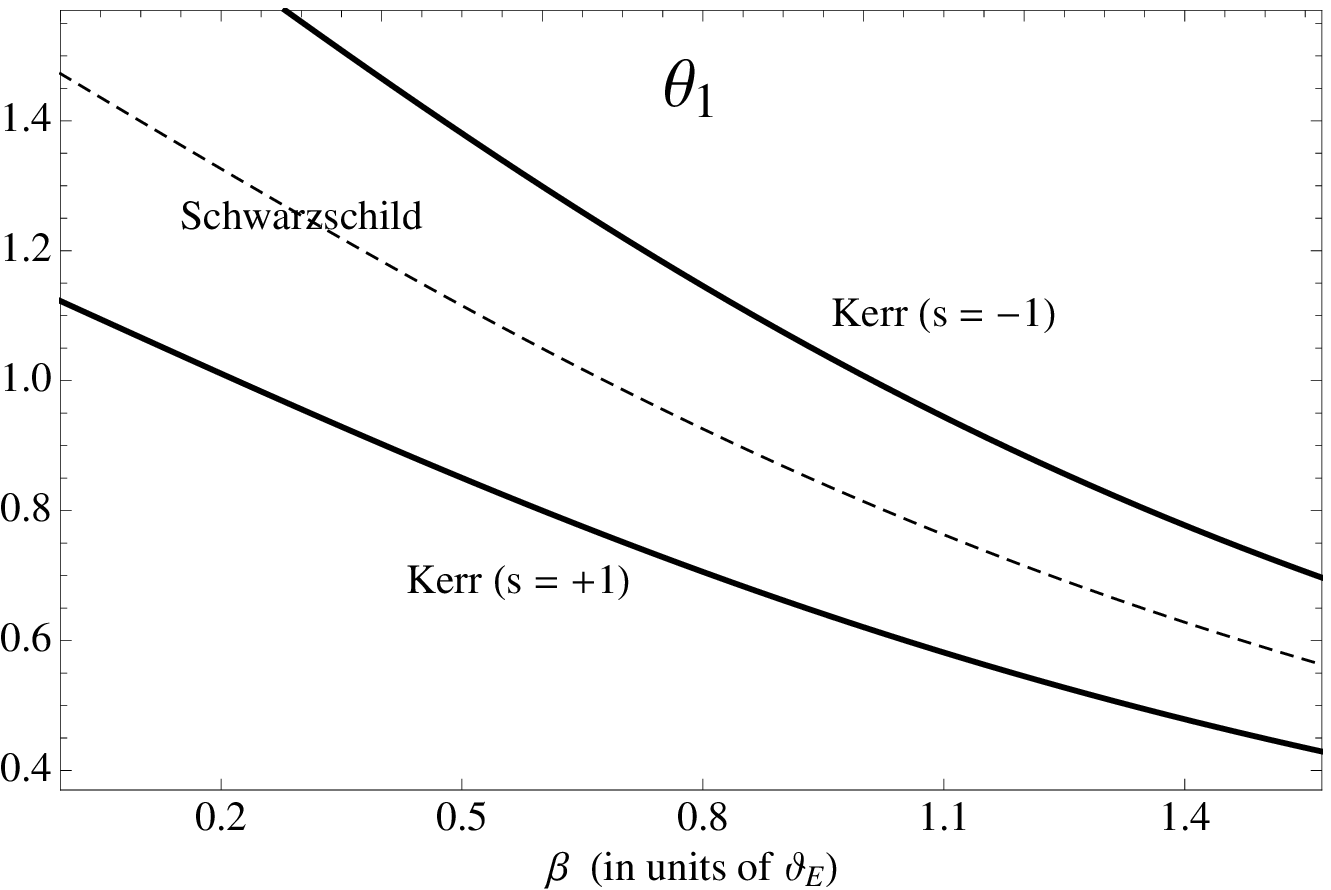}
\includegraphics[scale=.6]{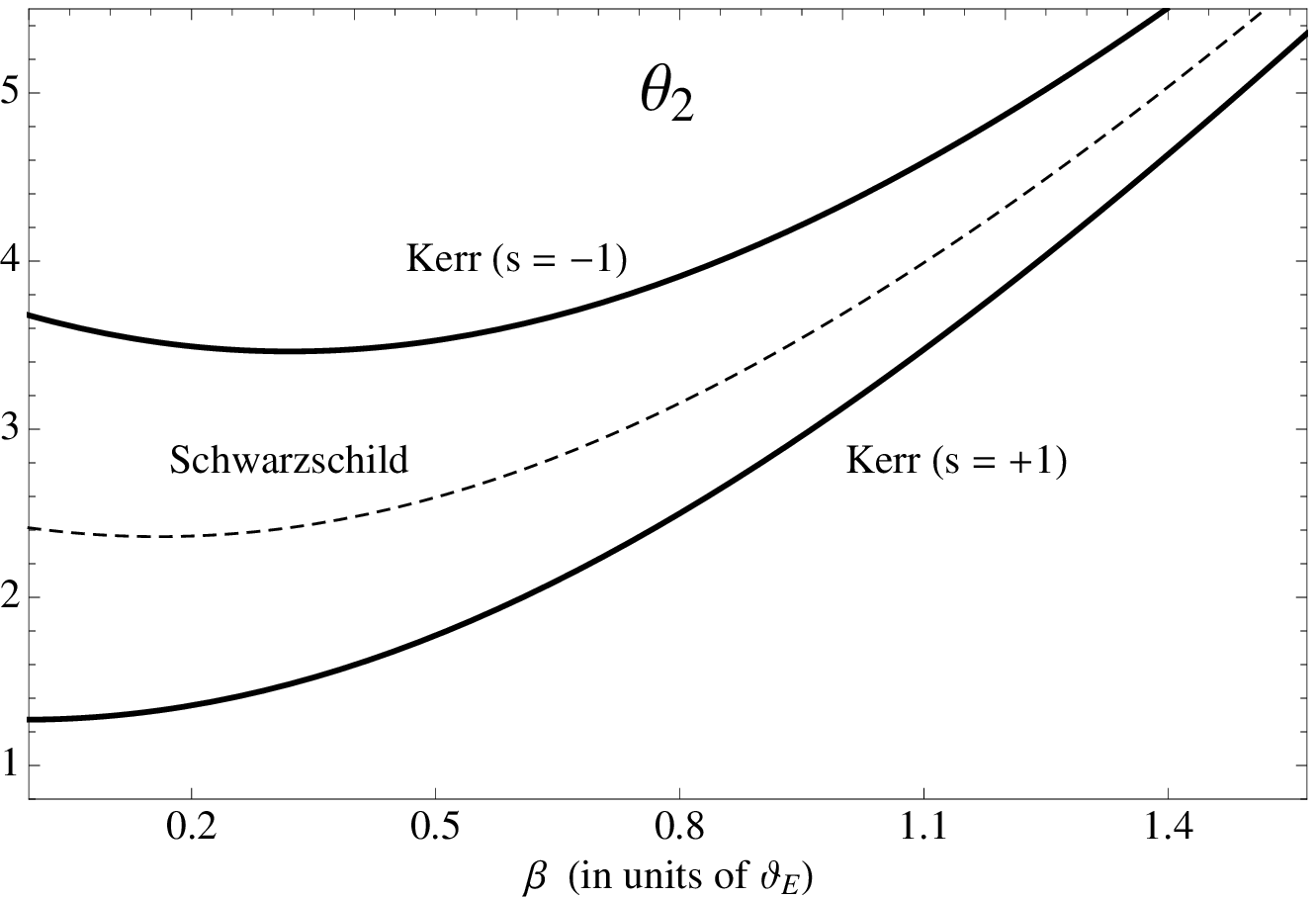}
\includegraphics[scale=.6]{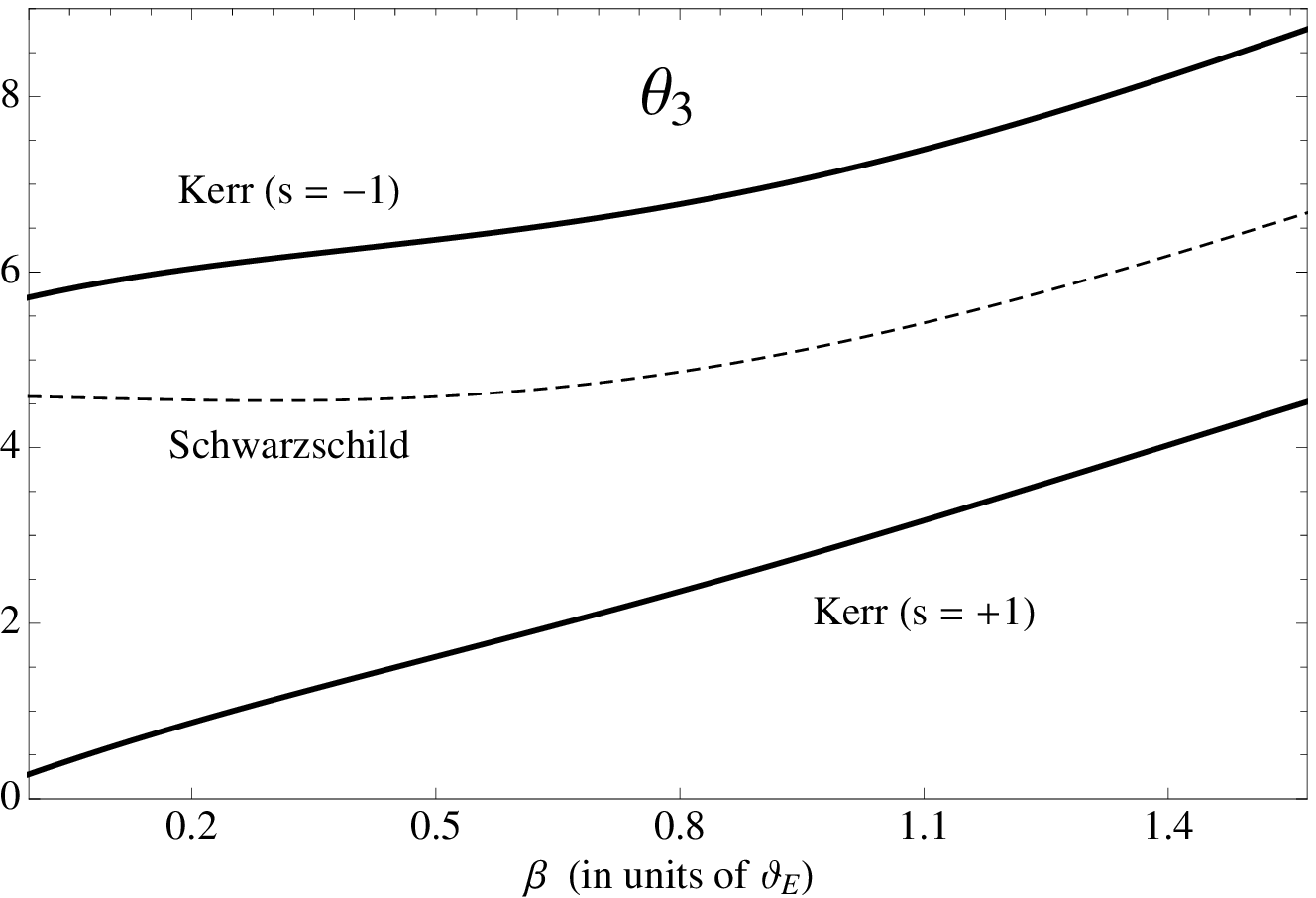}
\end{center}
\caption{
First-, second-, and third- order angular image correction terms as functions of the angular source position $\beta$, for a positive-parity image undergoing either prograde $(\sns = +1)$ or retrograde $(\sns = -1)$ motion near the equatorial plane of a Kerr black hole.  The solid curves represent a Kerr black hole with spin parameter $\ahat = 0.7$.  When $\ahat = 0$, we recover Schwarzschild lensing (dashed curves).  For the second- and third- order image corrections, the displacement parameter $\xi = 1$ and $D = d_{LS}/d_S = 0.5$.  (Note that $\theta_1, \theta_2$, and $\theta_3$ are dimensionless, but have factors of $\vep, \vep^2$, and $\vep^3$, respectively.  Note also that $d_{LS}$ and $d_S$ are the perpendicular distances between the lens and source planes and the observer and source plane, respectively.)  These results hold for a black hole with sufficiently small $\vth_E$.}
\label{fig:imagesource}
\end{figure}

In terms of the source position $\beta$, we can write the
terms for the positive- and negative-parity images as
\beqa
  \theta_0^{\pm} &=& \frac{1}{2} \left( \sqrt{\beta^2+4} \pm |\beta| \right) ,\nonumber\\
  \theta_1^{\pm} &=& \left( 1 \mp \frac{|\beta|}{\sqrt{\beta^2+4}} \right)
    \frac{15\pi - 16\,\sns^{\pm}\,\ahat}{32}\ ,\nonumber
\eeqa
where we have written $\sns^{\pm}$ to remind ourselves that the
two images have different respective values of the prograde/retrograde
sign parameter.  In fact, we have $\sns^- = -\sns^+$.  The terms $\theta_2^{\pm}$ and $\theta_3^{\pm}$ as functions of $\beta$ are similarly obtained, but are too lengthy to be written here.  Now thinking
of the universal relations studied in \cite{KP2}, we observe that
the zeroth-order terms obey
\beqa
  \theta_0^+ \ - \ \theta_0^- = |\beta|, \qquad
  \theta_0^+\,\theta_0^- = 1,\nonumber
\eeqa
which are identical to the zeroth-order position relations
obeyed by PPN models (see \cite{KP2}).  The first-order terms have
\beqa
  \theta_1^+ \ + \ \theta_1^- = \frac{15\pi}{16}
    \ + \ \frac{\sns^+\,\ahat\,|\beta|}{\sqrt{\beta^2+4}}\ .\nonumber
\eeqa
In \cite{KP2} it was shown that $\theta_1^+ +\, \theta_1^-$ is
independent of source position for static, spherical black
holes in all theories of gravity that can be expressed in
the PPN framework.
However, as first shown in \cite{WernerPetters}, we see that in the presence
of spin $\theta_1^+ + \theta_1^-$ is no
longer independent of source position.  This is a direct
consequence of the fact that one image corresponds to a
light ray that follows prograde motion, while the other
has retrograde motion.
The difference between the second-order components is (cf. \cite{KP2})
\beqa
  \theta_2^+ \ - \ \theta_2^- = &-&\!\!2\ahat^2\sqrt{4+\beta^2}~+~\frac{\ahat|\beta|(16\ahat + 15\pi\sns^{+})\sqrt{4+\beta^2}}{8(4 + \beta^2)^{3/2}}\nonumber\\
  &+&\frac{-30\pi\sns^{+}\ahat+\ahat(48\ahat+15\pi\sns^{+})\left(4+\beta^2\right)}{8\left(4+\beta^2\right)^{3/2}}~+~\frac{|\beta|\,{\tt I}}{256},\nonumber
\eeqa
where $${\tt I} = -4096 + 225\pi^2 + 2048 D^2 + 160\pi\sns^{+}\ahat - 512\ahat^2+ 
  4096D(1-D)\xi.$$
We can likewise consider $\theta_3^+ \pm\, \theta_3^-$, but will forgo writing them here.  Plots of the image correction terms $\theta_1, \theta_2$ and $\theta_3$ as functions of the source position $\beta$ are shown in \reffig{imagesource}, for a positive-parity image undergoing either prograde $(\sns = +1)$ or retrograde $(\sns = -1)$ motion. 

\subsection{Magnifications}
\label{sec:mag}

In Paper I we derived the following general magnification formula:
\beqa
\label{eq:mag}
  \mu = \left[ \frac{\sin\cb}{\sin\vth} \left(
      \frac{\partial\cb}{\partial\vth }\ \frac{\partial\chi}{\partial\vphi}
    - \frac{\partial\cb}{\partial\vphi}\ \frac{\partial\chi}{\partial\vth }
  \right) \right]^{-1} .
\eeqa
To compute $\partial\cb/\partial\vth$, we employ the same techniques that led to eqn.~(\ref{eq:lens3h}).
For $\partial\chi/\partial\vphi$, we use eqn.~(\ref{eq:lens4v}).
(Note that
$\partial\cb/\partial\vphi=0$ for quasi-equatorial lensing.)  The result is the following series expansion:
\beqa 
\label{eq:muser}
  \mu = \mu_0 \ + \ \mu_1\,\vep \ + \ \mu_2\,\vep^2 \ + \ \mu_3\,\vep^3
    \ + \ \order{\vep}{4} ,
\eeqa
where
\beqa
  \mu_0 &=& \frac{\theta_0^4}{\theta_0^4-1}\ ,\label{eq:mag0}\\
  \mu_1 &=& - \frac{[15\pi(\theta_0^2-1)^2 + 64\,\sns\,\ahat\,\theta_0^2]\,
    \theta_0^3} {16\,(\theta_0^2-1)^2 (\theta_0^2+1)^3}\ ,\label{eq:mag1}\\
\mu_2 &=& \frac{\theta_0^2}{384(\theta_0^2-1)^3(\theta_0^2+1)^5}\Bigg[768 \,\ahat^2 \theta_0^4 (5 - 2 \theta_0^2 + 5 \theta_0^4)\label{eq:mag2}\\
&&+~ 
   120\pi \sns^{+}\ahat (1 + 16 \theta_0^2 - 34 \theta_0^4 + 
      44 \theta_0^6 - 39 \theta_0^8 + 
      12 \theta_0^{10})\nonumber\\
       &&+~ \theta_0^2(\theta_0^2-1)^2\left(-12288 D (\theta_0 + \
\theta_0^3)^2 + 
      1024 D^2 (1 + \theta_0^2)^2 ( \theta_0^4 + 16 \theta_0^2+ 1)\right.\nonumber\\ 
      &&-~3\!\left.\theta_0^2 \left(4096 + \theta_0^2(-675 \pi^2 + 
            4096 (\theta_0^2+2))\right)\right) - 
   6144 (D-1) \left(\theta_0 - \theta_0^5)^2 (-1 + 
      2 D \theta_0^2 (2 + \theta_0^2)\right) \xi\Bigg]\ .\nonumber
      \label{eq:mag3}
\eeqa
Note that displacement terms (indicated by $\xi$) begin to appear only at second order in $\vep$.  The third-order magnification term $\mu_3$ is given in Appendix~\ref{app:third-order}.
In terms of the source position
$\beta$, we can write the terms for the positive- and negative-parity
images as
\beqa
  \mu_0^{\pm} &=& \frac{1}{2}
    \pm \frac{\beta^2+2}{2\,|\beta| \sqrt{\beta^2+4}}\ ,\nonumber \\
  \mu_1^{\pm} &=& - \frac{15\pi\,\beta^2 + 64\,\sns^\pm\,\ahat}
    {16\,\beta^2\,(\beta^2+4)^{3/2}}\ , \nonumber\\
  \mu_2^{\pm} &=&
    \pm \ \frac{2025\pi^2 - 1024 (\beta^2+4) (12+D(12-(\beta^2+18)D))}
      {384\,|\beta|\,(\beta^2+4)^{5/2}}
    \nonumber\\
  && + \ \frac{5\pi\,\sns^\pm\,\ahat}{32\,\beta^2} \left[ -1 \pm
      \frac{|\beta|\,(\beta^4+34\beta^2+48)}{(\beta^2+4)^{5/2}} \right]
  \ \pm \ 2\,\ahat^2\,\frac{5\beta^2+8}{|\beta|^3 (\beta^2+4)^{5/2}}
  \nonumber\\
  && + \ 16\,\xi\,(1-D)\left[ D \pm \frac{D(\beta^4+6\beta^2+6)-1}
      {|\beta|\,(\beta^2+4)^{3/2}} \right].\nonumber
\eeqa
Observe that
\beqa
  \mu_0^+ \ + \ \mu_0^- &=& 1\,,\nonumber \\
  \mu_0^+ \ - \ \mu_0^- &=& \frac{\beta^2+2}{|\beta|\,(\beta^2+4)^{1/2}}\ ,\nonumber \\
  \mu_1^+ \ + \ \mu_1^- &=& - \frac{15\pi}{8(\beta^2+4)^{3/2}}\ ,\nonumber \\
  \mu_1^+ \ - \ \mu_1^- &=& - \frac{8\,\sns^+\,\ahat}
    {\beta^2 (\beta^2+4)^{3/2}}\ ,\nonumber \\
  \mu_2^+ \ + \ \mu_2^- &=& \frac{5\pi\,\sns^{+}\,\ahat}{16}\frac{|\beta|\,(\beta^4+34\beta^2+48)}{\beta^2(\beta^2+4)^{5/2}}\ +\  32\,\xi\,D (1-D)\ ,\nonumber\\
    \mu_2^+ \ - \ \mu_2^- &=& - \ \frac{5\pi\,\sns^{+}\,\ahat}{16\,\beta^2}
  \ + \ 4\,\ahat^2\,\frac{5\beta^2+8}{|\beta|^3 (\beta^2+4)^{5/2}}\ + \ 32\,\xi\,(1-D)\left[\frac{D(\beta^4+6\beta^2+6)-1}
      {|\beta|\,(\beta^2+4)^{3/2}} \right]\nonumber\\
      &&+\ \frac{2025\pi^2 - 1024 (\beta^2+4) (12+D(12-(\beta^2+18)D))}
      {192\,|\beta|\,(\beta^2+4)^{5/2}}\ .\label{eq:mu2}
\eeqa      
The expressions $\mu_3^+ \pm \mu_3^-$ are given in Appendix~\ref{app:third-order}.  The zeroth-order sum relation is the same as the universal
relation found for static, spherical PPN models in \cite{KP2}.
Notice that the zeroth-order difference relation is independent
of spin.  In the first-order difference relation, the right-hand
side is zero for PPN models, but nonzero in the presence of spin (see also \cite{WernerPetters}).
In the second-order sum relation, the
right-hand side is not zero even in the absence of spin.  This is a consequence of the displacement (indicated by $\xi$).

\subsection{Critical and Caustic Points}
\label{sec:caustics}
To determine the set of critical points, we set $\mu^{-1} = 0$, the reciprocal of the series expansion given by eqn.~(\ref{eq:muser}) in Section~\ref{sec:mag}, and solve for $\theta_0, \theta_1, \theta_2$, and $\theta_3$.  This yields the following $\theta$-components:
\beqa
{\theta_0}_{\tt critical}^{\pm} &=& 1\ ,\label{crit1}\\
{\theta_1}_{\tt critical}^{\pm} &=& -\sns\,\ahat + \frac{15\pi}{32}\ ,\label{crit2}\\
{\theta_2}_{\tt critical}^{\pm} &=& 8 - \frac{15\pi\,\sns\,\ahat}{32} - \frac{675\pi^2}{2048} + D^2\left(\frac{20}{3} - 8\,\xi\right) + 4D(3\,\xi-2)  - 4\xi\ ,\label{crit3}\\
{\theta_3}_{\tt critical}^{\pm} &=& -\sns\,\ahat\,\left(\frac{225\pi^2}{256} - 8D^2\,(1-\xi) - 8D\,\xi - 8\right) -\frac{15\pi\,\ahat^2}{64}\nonumber\\
&&~- \frac{15\pi (400 - 225\pi^2 - 4096 D^2 (1 - \xi) + 2048\,\xi - 
   2048 D (-2 + 3\,\xi))}{8192}\ ,\label{crit3}
\eeqa
where $``\pm"$ corresponds to the two values $\sns = \pm 1$.  Note that since we are in the regime of quasi-equatorial Kerr lensing ($\vphi = \vphi_0 + \dvphi$ with $\vphi_0 = 0$ (retrograde motion) or $\pi$ (prograde motion)), eqns.~(\ref{crit1})--(\ref{crit3}) do not define a circle on the lens plane, but are to be interpreted (by eqn.~(\ref{eq:app:sdef})) as four points $(\theta_{\text{crit}}^{+},\pi \pm \dvphi), (\theta_{\text{crit}}^{-},\pm \dvphi)$ on the lens plane, for a given $\dvphi$.  We now insert these into the ``horizontal" lens equation (\ref{eq:lens3h}) to third order in $\vep$ and solve for $\beta$.  This yields the $\beta$-components of the caustic points, which we express here as a series expansion in $\vep$ to third order:
\beqa
\label{causticpts}
\beta_{\tt caustic}^{\pm} &=& -\sns\,\ahat\,\vep - \frac{5\pi\,\sns\,\ahat}{16}\,\vep^2\\
&&~+\frac{\ahat}{512}\Biggl[1136\pi\,\ahat + 
  \sns\left(225\pi^2 - 4096\,\xi + 4096 D^2 (1 - 2\xi) + 
     4096 D (-2 + 3\xi)\right)\Biggr]\,\vep^3\nonumber\\&&~+\order{\vep}{4}\ ,\nonumber
\eeqa
The signs $\pm$ correspond to prograde ($\sns = + 1$) and retrograde ($\sns = -1$) motion, respectively.  When $\ahat = 0$ the two caustic points converge to one point at the origin of the source plane.  Note from the third-order term that the caustic points are not symmetric about the vertical axis on the light source plane.

\subsection{Total Magnification and Centroid}
\label{sec:totmag}

If the two images are too close together to be resolved (as in
microlensing), the main observables are the total unsigned magnification
and the magnification-weighted centroid position.  Using our
results above, we compute the total unsigned magnification:
\beq \label{eq:abs-mag}
  \mu_{\rm tot} = |\mu^+|\ +\ |\mu^-|
  = \frac{\beta^2+2}{|\beta|\,(\beta^2+4)^{1/2}}
    \ - \ \frac{8\,\sns^+\,\ahat}{\beta^2 (\beta^2+4)^{3/2}}\ \vep \\
    \ + \ (\mu_2^{+}\, -\, \mu_2^{-})\ \vep^2\ +\ (\mu_3^{+}\, -\, \mu_3^{-})\ \vep^3\ +\ \order{\vep}{4} ,
\eeq
where the second- and third- order terms are given by eqns.~(\ref{eq:mu2}) and (\ref{eq:mu3}), respectively.  The magnification-weighted
centroid position (actually, its ``horizontal" component, since we are in the regime of quasi-equatorial lensing) is
\beqa \label{eq:hor-centroid}
  \Theta_{\rm cent} = \frac{\theta^{+} |\mu^{+}| - \theta^{-} |\mu^{-}|}
    {|\mu^{+}| + |\mu^{-}|}
  &=& \frac{|\beta|\,(\beta^2+3)}{\beta^2+2}
    \ + \ \frac{(2-\beta^2)\,\sns^+\,\ahat}{(\beta^2+2)^2}\ \vep\nonumber\\
    &&~+ \ \frac{(4+\beta^2)^2\,{\tt C_{2,1}}}{384\,|\beta|\,(8+6\beta^2+\beta^4)^3}\  \vep^2\ +\  \Theta_{{\rm cent},3}\ \vep^3\ +\ \order{\vep}{4}\ ,
    \eeqa
where
 \beqa
{\tt C_{2,1}} &=& 120\pi\,\sns^{+}\,\ahat\,|\beta|\,(2 + \beta^2) (3 + \beta^2) (4 +
\beta^2)^{3/2} + 
  384\,\ahat^2 \left[(2 + \beta^2) (-16 - 8 \beta^2 + \beta^4)\right.\nonumber\\ 
  &&\left.~+\  4 (8 + 2 \beta^2 + \beta^4)\right] - \beta^2 (2 + \beta^2) \left[3
(675\pi^2 - 4096 (4 + \beta^2))\right.\nonumber\\ 
&&\left.~+\ 1024 (4 + \beta^2) \left(D (6 \beta^2 - 
           D (-2 + 9 \beta^2 + \beta^4)) + 
        3 (-1 + D) (-\beta^2 + 
           2 D (6 + 4 \beta^2 + \beta^4))\,\xi\right)\right]\ ,\nonumber
           \eeqa
and where the third-order term $\Theta_{{\rm cent},3}$ is given in Appendix~\ref{app:third-order}.  In \cite{KP2} it was shown that the first-order corrections to the
total unsigned magnification and centroid position vanish universally
for static, spherical black holes that can be described in
the PPN framework.  In the presence of spin, the first-order
corrections are nonzero.  Once again, the displacement terms (indicated by $\xi$) appear
only at second order in $\vep$ in both eqns.~(\ref{eq:abs-mag}) and (\ref{eq:hor-centroid}).

\subsection{Time Delay}
\label{sec:Tdel}

In \refapp{Kerr-Tdel}, we show that the lensing time delay can
be written as
\beqa \label{eq:tdel}
  c \tau = T(R_{\rm src}) + T(R_{\rm obs}) - \frac{d_S}{\cos\cb}\ ,\nonumber
\eeqa
where
\beqa \label{eq:Rdef}
  R_{\rm obs} = d_L\,, \qquad
  R_{\rm src} = \left( d_{LS}^2 + d_S^2 \tan^2 \cb \right)^{1/2}\,,\quad
  \cb = 4\,\beta\,D\,\vep\ ,\nonumber
\eeqa
and $R_{\rm obs}$ and $R_{\rm src}$ are the radial coordinates of the observer and source in the Kerr
metric.
We derive a Taylor
series expansion for the function $T(R)$ in \refapp{Kerr-Tdel}
(see eqn.~(\ref{eq:app:Tser})).
To determine the observable time delay, we evaluate $T(R)$ at
$R_{\rm src}$ and $R_{\rm obs}$, and then replace $r_0$ with
$b$ using eqn.~(\ref{eq:rofb}).  We change to angular variables using
$b = d_L \sin\vth$, and then reintroduce the scaled angular
variables in eqns.~(\ref{eq:newvar})--(\ref{eq:tseries}).  Finally, we take a formal Taylor
series to second order in our expansion parameter $\vep$.  This yields
\beqa \label{eq:app:tau}
  \frac{\tau}{\tau_E} =
    \frac{1}{2} \left[ 1 + \beta^2 - \theta_0^2 - \ln \left(
    \frac{d_L\,\theta_0^2\,\vthE^2}{ 4\,d_{LS}} \right) \right]
  \ + \ \frac{15\pi - 16\,\sns\,\ahat}{16\,\theta_0}\ \vep\ 
  +\ \frac{{\tt T}}{1536\,\theta_0^2\,(\theta_0^2 + 1)}\ \vep^2 \ +\  \order{\vep}{3}\ ,\nonumber\\
\eeqa
where 
\beqa
{\tt T} &=&-96\pi\,\sns\,\ahat\,(-7 + \theta_0^2 + \theta_0^4) + 768\,\ahat^2\,(2 \theta_0^2 + 3 \theta_0^4 + \theta_0^6) - (1 + \theta_0^2) \left\{675 \pi^2\right.\nonumber\\
&&~+\left.3072\,\theta_0^2\,(\theta_0^2 + 1) \left(2+\beta^4+\theta_04-2\beta^2\,(\theta_0^2+1)-4\xi\right)\right.\nonumber\\
    &&~+\left.3072 D \theta_0^2 (1 + \theta_0^2) (-8 + \beta^4 + 
     2 \beta^2 \theta_0^2 - 3 \theta_0^4 + 4 \xi + 
     8 \theta_0^2 \xi)\right.\nonumber\\
      &&~+\left.1024 D^2 (1 + \theta_0^2) \left(-8 + (24 - 5 \beta^4) \theta_0^2 + 
     5 \theta_0^6 - 24 \theta_0^4 \xi\right)\right\}\label{T2}\ ,
\eeqa
and the natural lensing time scale is
\beqa \label{eq:app:tauE}
  \tau_E \equiv \frac{d_L d_S}{c\,d_{LS}}\,\vthE^2
  = 4\,\frac{\gravr}{c}\ .\nonumber
\eeqa
Notice that retrograde motion ($\sns = -1$) leads to a longer
time delay than prograde motion ($\sns = +1$), which makes sense
intuitively.  As with our other lensing observables, displacement terms in the time delay (indicated by $\xi$) begin to appear only at second order in $\vep$.

The differential time delay between the two images,
$\Delta\tau = \tau^- - \tau^+$ is such that
\beqa
\frac{\Delta\tau}{\tau_E} & =  & \left[ \frac{1}{2}\,|\beta|\,\sqrt{\beta^2+4}
+ \ln\left(\frac{\sqrt{\beta^2+4}+|\beta|}{\sqrt{\beta^2+4}-|\beta|}\right)
\right] 
\ + \ \left[ \frac{15\pi}{16}\,|\beta|
  + \sns^+\,\ahat \sqrt{\beta^2+4} \right] \vep\nonumber\\
  &&~+ \ \frac{{\tt D}}{1536(\beta^2+4)}\ \vep^2 + \order{\vep}{3}\ ,\label{eq:timedelay}
\eeqa
where
\beqa
{\tt D} &=& 96\pi\,\sns^{+}\,\ahat\,(4 + \beta^2)(7\beta^2-1) - 768\,\ahat^2\,|\beta| \sqrt{4 + \beta^2} - |\beta| \sqrt{
 4 + \beta^2}\,\left\{-675 \pi^2 (3 + \beta^2)\right.\nonumber\\
 &&~+\left. 
      3072 (-8 + 2\beta^2 + \beta^4) +1024 D^2 (4 + \beta^2) (18 + 5 \beta^2 - 24 \xi) - 
   3072 D (4 + \beta^2) (6 + \beta^2 - 8 \xi)\right\}.\nonumber\\\label{D2} 
\eeqa

\section{Remarks on Lensing Observables}

We make a few remarks regarding our results:

\begin{enumerate}

\item The procedure for solving the lens equations in the quasi-equatorial regime is as follows: given a source whose (scaled) location on the source plane is $(\beta,\chi_0 + \dchi)$, we first solve the ``horizontal" lens equation (\ref{eq:lens3h}) term by term to find $\theta_0, \theta_1, \theta_2$, and $\theta_3$ (all expressed in terms of $\beta$), and then insert these into the ``vertical" lens equation (\ref{eq:lens4v}) and solve for $\dvphi$.  The (scaled) locations of the two images in the lens plane are then
$$
\left(\theta_0 + \theta_1^{\pm}\,\vep + \theta_2^{\pm}\,\vep^2 + \theta_3^{\pm}\,\vep^3~,~\vphi_0 + \dvphi^{\pm}\right)\ ,
$$ 
where $``\pm"$ corresponds to $\sns = \pm 1$ and where $\vphi_0 = 0$ for retrograde motion $(\sns = -1)$ and $\pi$ for prograde motion $(\sns = +1)$.

\item Note that for all lensing observables---image position, image magnification, total unsigned magnification, centroid, and time delay---the displacement parameter $\xi$ begins to appear only at second order in $\vep$.  {\it Therefore displacement can safely be ignored  for studies of first-order corrections to weak-deflection quasi-equatorial Kerr lensing.}  Note that the displacement affects the caustic positions only at third order in $\vep$.

\item When there is no spin, we obtain new results on the lensing observables due to Schwarzschild lensing with displacement.
Indeed, all of our results in Section~\ref{sec:lensing} (including the third-order results in Appendix~\ref{app:third-order}) immediately apply to this regime once we set $\ahat = 0$ and the displacement parameter $\xi = 1$.  This is equivalent to beginning with the spherically symmetric lens equation with displacement (given in Bozza \& Sereno \cite{BS} and eqn.~(19) in Paper I) and then computing lensing observables perturbatively in $\vep$.

\item If one sets $\ahat = 0 =\xi$ (i.e., if one turns off spin and ignores displacement), then all of our results in Section~\ref{sec:lensing} are consistent with the previous studies of Keeton \& Petters \cite{KP1,KP2}.

\item The total magnification and centroid (eqns.~(\ref{eq:abs-mag}) and (\ref{eq:hor-centroid}), respectively)  are consistent with the corresponding results obtained in Werner \& Petters \cite{WernerPetters} to first order in $\vep$.  (The analysis in \cite{WernerPetters} was carried to first order in $\vep$ and did not consider displacement.)  In fact we point out that the ``horizontal" and  ``vertical" components of our lens equation (eqns.~(\ref{eq:lens3h}) and (\ref{eq:lens3v})) and our magnification terms (eqns.~(\ref{eq:mag0}) and (\ref{eq:mag1})) are all consistent to first order in $\vep$ with those in \cite{WernerPetters}, after an appropriate change of variables.  Furthermore, the ``horizontal" and ``vertical" components of our bending angle (see eqns.~(\ref{eq:horbangle}) and (\ref{eq:bavy}) in Appendix~\ref{app:Kerr-bend-angle} below) are also consistent to second order in $\vep$ with those in \cite{WernerPetters} (their bending angles were written to second order in $\vep$).

\item Finally, our image correction and magnification terms are also consistent with those in Sereno \& De Luca \cite{SerenoDeLuca} to first order in $\vep$, while the ``horizontal" and ``vertical" components of our bending angle are consistent to second order in $\vep$.

\end{enumerate}

\section{Conclusions}
\label{sec:conclusions}

In this paper we derived analytical expressions for the lensing observables for the case of quasi-equatorial lensing with displacement by a Kerr black hole.  We calculated the  light bending angle explicitly
for an equatorial observer and light rays that are quasi-equatorial, and then applied this to a perturbative
framework to third order in the invariant parameter 
$\vep$, which is the ratio of
the angular gravitational radius to the angular Einstein radius.
We obtained new formulas and results for the fundamental lensing observables: 
image position, image magnification, total unsigned
magnification, centroid, all to third order in $\vep$, and time delay to second order.  Our results made explicit the effect of the displacement that occurs when the tangent lines to the rays from the source and observer do not intersect on the lens plane.  We also showed that displacement effects begin to appear at second order in $\vep$, and so can safely be ignored for
studies of first-order corrections to weak-deflection quasi-equatorial
Kerr lensing.  Our findings also yield new analytical results for Schwarzschild lensing with displacement.

This analysis has allowed us to go beyond previous work and probe deeper into the gravitational field of a Kerr black hole, by providing explicit perturbative analytical formulas showing how each lensing osbervable is affected by higher-order terms.  Our results should be useful in observing general relativistic corrections, and can also be used as a tool in testing Einstein's theory and perhaps also Cosmic Censorship.

\begin{acknowledgments}

ABA and AOP would especially like to thank Marcus C. Werner for helpful discussions.  AOP acknowledges the support of NSF Grant DMS-0707003.

\end{acknowledgments}

\appendix
\section{Transformation from Sky Coordinates to Boyer-Lindquist Coordinates}
\label{app:sec:BHCoords}

\begin{figure}[t]
\begin{center}
\includegraphics[scale=.62]{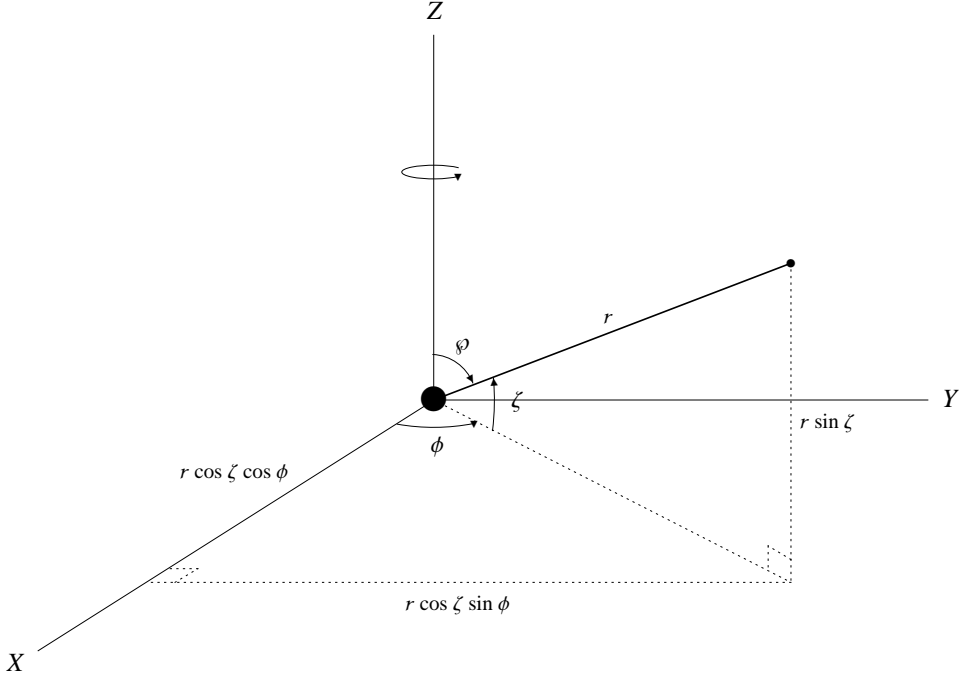}
\end{center}
\caption{
Cartesian $(X,Y,Z)$ and spherical polar $(r,\kpolar,\kazym)$
coordinates centered on the black hole, where $\kpolar = \pi/2 - \wp$ with $\wp$ the polar angle.  The black hole spins about the $Z$-axis, which corresponds to
$\kpolar=\pi/2$, in the direction of
increasing $\kazym$.  The equatorial plane
of the black hole corresponds to $\kpolar = 0$ or the
$(X,Y)$-plane.  Taken from Fig.~4 in Paper I \cite{AKP}.}
\label{fig:BHCoords}
\end{figure}

In this section we determine the relation between angular coordinates
$(\vth, \vphi)$ on the sky as measured by the observer, and the slightly
modified Boyer-Lindquist coordinates $(t,r,\kpolar,\kazym)$ shown in \reffig{BHCoords}.

\begin{figure}[t]
\begin{center}
\includegraphics[scale=.62]{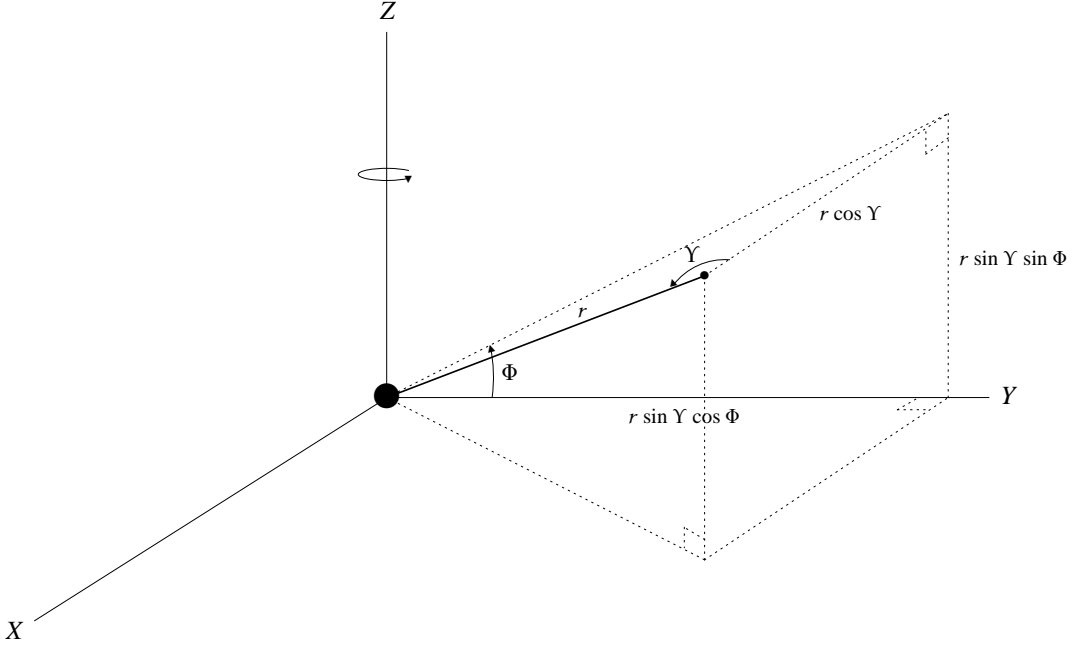}
\end{center}
\caption{
Lens-centered coordinates $(r,\Upsilon, \Phi)$.}
\label{fig:app:LensCentered}
\end{figure}

Recall from Paper I that 
the latter coincide with
the usual Boyer-Lindquist coordinates $(t,r,\wp,\kazym)$, except that
the polar angle $\wp$ is shifted to
$\kpolar = \pi/2-\wp$.
To analyze light bending, it is actually convenient to work with another set of coordinates, namely, the lens-centered
coordinates $(r,\Upsilon, \Phi)$ shown in \reffig{app:LensCentered}. 
Our goal is to connect the modified 
Boyer-Lindquist coordinate angles $(\kpolar,\kazym)$ to 
observer-centered angles $(\vth, \vphi)$.  This will be done
in two stages: first, by relating
$(\kpolar,\kazym)$ to 
$(\Upsilon, \Phi)$, and then by relating $(\Upsilon, \Phi)$ to $(\vth,\vphi)$.  This simplifies the ray-tracing and the resulting geodesic equations.

Comparing \reffig{BHCoords} with Fig.~\ref{fig:app:LensCentered} below yields the following relation between $(\kpolar,\kazym)$ and 
$(\Upsilon, \Phi)$:
\beq \label{eq:app:tp2ex}
  \sin\kpolar = \sin\Upsilon \sin\Phi\,, \qquad
  \tan\kazym = \tan\Upsilon \cos\Phi\,.
\eeq
In order to relate the observer-centered angles $(\vth,\vphi)$
to the lens-centered angles $(\Upsilon,\Phi)$, we make the following
construction.  Consider extending the actual light ray to infinity
both ``behind'' the source and ``beyond'' the observer.  Note that evaluating at such points is well-defined because the light ray is a linear path in the asymptotically flat regions where both the source and observer reside.  With that said, the
asymptotic ``final'' angular position the light ray reaches is (cf. Section~\ref{sec:bend-angle})
\beqa
  \Upsilon_f &=& \vth\,, \qquad \Phi_f = \vphi+\pi \hspace{.5in}\mbox{(prograde motion),}\nonumber\\
  \Upsilon_f &=& -\vth\,, \qquad \Phi_f = \vphi \hspace{.56in}\mbox{(retrograde motion).}\nonumber
\eeqa
The asymptotic ``initial'' angular position from which the light
ray originates is
\beqa
  \Upsilon_i &=& \pi-\vth_S\,, \qquad \Phi_f = \vphi_S \hspace{.6in}\mbox{(prograde motion),}\nonumber\\
  \Upsilon_i &=& \pi+\vth_S\,, \qquad \Phi_f = \vphi_S \hspace{.5in}\mbox{(retrograde motion).}\nonumber
\eeqa
Using eqn.~(\ref{eq:app:tp2ex}), we can find the initial and final positions
in terms of the angles $(\kazym,\kpolar)$.  For prograde motion, these are:
\beqa \label{eq:app:tp2ex2}
  \sin\kpolar_i =  \sin\vth_S \sin\vphi_S\,, &&\qquad
  \sin\kpolar_f = -\sin\vth \sin\vphi\,,\label{eq:app:ToEtaXi}\\
  \tan\kazym_i  = -\tan\vth_S \cos\vphi_S\,, &&\qquad
  \tan\kazym_f  = -\tan\vth \cos\vphi\,.\nonumber
\eeqa
For retrograde motion, they are:
\beqa \label{eq:app:ex2tp2}
   \sin\kpolar_i = -\sin\vth_S \sin\vphi_S\,, &&\qquad
  \sin\kpolar_f = \sin\vth \sin\vphi\,,\label{eq:app:FromEtaXi}\\
  \tan\kazym_i  = \tan\vth_S \cos\vphi_S\,, &&\qquad
  \tan\kazym_f  = \tan\vth \cos\vphi\,.\nonumber
\eeqa
We will use eqns.~(\ref{eq:app:ToEtaXi}) and (\ref{eq:app:FromEtaXi}) in our derivation of the ``vertical" component of the bending angle vector in Appendix~\ref{sec:app:vertical} below.

\section{Quasi-Equatorial Kerr Bending Angle}
\label{app:Kerr-bend-angle}

\subsection{Equations of Motion for Quasi-Equatorial Null Geodesics}

Recall from Appendix~A\,1 of Paper I that the equations of motion for null geodesics are
\beqa
  \hat{\dot{t}} &=& 1 + \frac{2\gravr\,r(\bha^2-\bha\hcL+r^2)}
    {[\bha^2+r(r-2\gravr)] (r^2 + \bha^2 \sin^2\kpolar)}\ ,
    \label{eq:app:tdot} \\
  \hat{\dot{r}} &=& \pm \frac{[ r^4 - (\hcQ + \hcL^2-\bha^2) r^2
    + 2\gravr ((\hcL - \bha)^2+ \hcQ) r - \bha^2 \hcQ ]^{1/2}}
    {r^2 + \bha^2 \sin^2\kpolar}\ ,
    \label{eq:app:rdot} \\
  \hat{\dot{\kazym}} &=& \frac{ 2 \bha \gravr r
    + \hcL r (r-2\gravr) \sec^2\kpolar + \bha^2 \hcL \tan^2\kpolar }
    {[\bha^2+r(r-2\gravr)] (r^2 + \bha^2 \sin^2\kpolar)}\ .
    \label{eq:app:xdot}\\
     \hat{\dot{\kpolar}} &=& \pm \frac{( \hcQ + \bha^2 \sin^2\kpolar
    - \hcL^2 \tan^2\kpolar )^{1/2}}{r^2 + \bha^2 \sin^2\kpolar}\ ,
    \label{eq:app:edot}
\eeqa
where $\hcQ = \mathcal{Q}/\mathcal{E}^2$ and $\hcL = \mathcal{L}/\mathcal{E}$, with $\mathcal{E}$ the energy, $\mathcal{L}$ the orbital angular momentum, and $\mathcal{Q}$ the Carter constant \big($\bha$ and $\gravr$ are given by eqn.~(\ref{eq:agravr})\big).  Now consider an equatorial observer and source in the asymptotically
flat region.  To compute the light bending angle, we focus on null geodesics
that remain close to the equatorial plane (which is a plane
of reflection symmetry).  There are light rays in the equatorial
plane that have $\kpolar = 0$ everywhere.  There are other light
rays that remain close to the plane and have $|\kpolar| \ll 1$
everywhere.  Such quasi-equatorial light rays must have
$\vphi = \vphi_0 + \delta\vphi$ with $\vphi_0$ equal to either
$0$ or $\pi$, and $|\dvphi| \ll 1$.  Given the spin configuration,
light rays with $\vphi_0 = 0$ follow retrograde motion and have
$\cL < 0$, while light rays with $\vphi_0 = \pi$ follow prograde
motion and have $\cL > 0$.  Thus, if we define a sign $\sns$ by
\beqa
\label{sns}
  \sns = {\rm sign}(\cL) = \cases{
    +1 & prograde motion, \cr
    -1 & retrograde motion,
  }
\eeqa
then we can identify
\beqa
\label{eq:app:sdef}
  \sns = -\cos\vphi_0\,.
\eeqa
We showed in Paper I that the constants of the motion $\hcL$ and $\hcQ$ can be written as
$$
\hcL = -d_L\sin\vth\,\cos\vphi\,,\qquad\hcQ = d_L^2\sin^2\vth\,\sin^2\vphi\,.
$$
In the quasi-equatorial regime, these become
\beqa
  \hcL = \sns\,b\,\cos\dvphi\,, \qquad
  \hcQ = b^2\,\sin^2\dvphi\,.\nonumber
\eeqa
We expect $\dvphi$ to be of the same order as $\kpolar$, so we
can Taylor expand eqns.~(\ref{eq:app:tdot})--(\ref{eq:app:edot}) in both $\kpolar$ and
$\dvphi$.  This yields
\beqa
  \hat{\dot{t}} &=& \frac{r^2}{\bha^2+r(r-2\gravr)}
    \left( 1 + \frac{\bha^2}{r^2} - \frac{2\,\gravr\,\bha\,b\,\sns}{r^3}\,\ssf
    \right)
    \ + \ \order{2}{} , \label{eq:app:tdot2}\\
  \hat{\dot{r}} &=& \pm \left( 1 - \frac{b^2}{r^2}\,\ssg
    + \frac{2\,\gravr\,b^2}{r^3}\,\ssf^2 \right)^{1/2}
    \ + \ \order{2}{} , \label{eq:app:rdot2} \\
  \hat{\dot{\kazym}} &=& \frac{b\,\sns}{\bha^2+r(r-2\gravr)}
    \left( 1 - \frac{2\gravr}{r}\,\ssf \right)
    \ + \ \order{2}{} , \label{eq:app:xdot2} \\
  \hat{\dot{\kpolar}} &=& \pm \frac{b}{r^2} \left[(\dvphi)^2
     - \ssg\,\kpolar^2\right]^{1/2}
    \ + \ \order{2}{} , \label{eq:app:edot2}
\eeqa
where $\order{2}{}$ indicate terms that are second order in
$\kpolar$ and/or $\dvphi$, and we have defined
\beqa \label{eq:FG}
  \ssf \equiv 1 - \sns\,\frac{\bha}{b} = 1 - \sns\,\ahat\,\frac{\gravr}{b}\ , \qquad
  \ssg \equiv 1 - \frac{\bha^2}{b^2} = 1 - \ahat^2\,\frac{\gravr^2}{b^2}\ .\nonumber
\eeqa
Notice that $\dot{t}$, $\dot{r}$, and $\dot{\kazym}$ do not depend
on $\kpolar$ or $\dvphi$ at zeroth order or first order.  In other
words, the ``in plane'' motion is insensitive to small displacements
above or below the equatorial plane.  By contrast, $\dot{\kpolar}$
lacks a zeroth-order term but has a nonzero first-order term.
Thus, there is a solution with $\kpolar = \dvphi = 0$ (i.e., a ray
that stays in the equatorial plane), but there are also solutions
in which $\kpolar$ and $\dvphi$ are nonzero.

Before evaluating the quasi-equatorial light bending, we need to relate the
light ray's coordinate distance of closest approach, $r_0$, to
the invariant impact parameter $b$.  The distance of closest
approach is given by the solution of $\dot{r} = 0$.  From
eqn.~(\ref{eq:app:rdot2}) this is a simple quadratic equation in $b$,
whose positive real solution is
\beq \label{eq:app:bofr}
  \frac{b}{r_0} = \left( \ssg - \frac{2\gravr}{r_0}\,\ssf^2 \right)^{-1/2} .
\eeq
Alternatively, $\dot{r} = 0$ is a cubic equation in $r_0$,
whose one real solution is given by
\beq \label{eq:rofb}
  \frac{r_0}{b} = \frac{2}{3^{1/2}}\ \ssg^{1/2}\,
    \cos\left[\frac{1}{3} \cos^{-1}\left(- 3^{3/2}\,\frac{\ssf^2}{\ssg^{3/2}}\ 
    \frac{\gravr}{b} \right) \right] .
\eeq
Taylor expanding in $\gravr/b \ll 1$ yields
\beqa
  \frac{r_0}{b} &=& \ssg^{1/2} 
    \ - \ \frac{\ssf^2}{\ssg}                     \left(\frac{\gravr}{b}\right)
    \ - \ \frac{3\,\ssf^4}{2\,\ssg^{5/2}}         \left(\frac{\gravr}{b}\right)^2
    \ - \ \frac{4\,\ssf^6}{\ssg^4}                \left(\frac{\gravr}{b}\right)^3
    \ - \ \frac{105\,\ssf^8}{8\,\ssg^{11/2}}      \left(\frac{\gravr}{b}\right)^4 
    \nonumber\\
  &&
    \ - \ \frac{48\,\ssf^{10}}{\ssg^7}            \left(\frac{\gravr}{b}\right)^5
    \ - \ \frac{3003\,\ssf^{12}}{16\,\ssg^{17/2}} \left(\frac{\gravr}{b}\right)^6
    \ + \ \order{\frac{\gravr}{b}}{7} .
    \label{eq:app:rb-ser}
\eeqa
(We could further expand $\ssf$ and $\ssg$ as Taylor series in
$\gravr/b$, but choose not to do that yet.)  Note that in the
absence of spin ($\bha =0$), $\ssf = \ssg = 1$ and so
eqns.~(\ref{eq:app:bofr})--(\ref{eq:app:rb-ser}) reduce to their
respective Schwarzschild values in \cite{KP1}.

\subsection{``Horizontal" Light Bending Angle}
\label{sec:app:horizontal}

We consider the bending of a null geodesic along the
$\phi$-direction (horizontal).  From
eqns.~(\ref{eq:app:rdot2}) and (\ref{eq:app:xdot2}), we can write the equation
of motion as
\beq \label{eq:app:hintgnd}
  \frac{d\kazym}{dr} = \frac{\hat{\dot{\kazym}}}{\hat{\dot{r}}}
  = \pm \frac{\sns\,b\,r^{1/2}\,(r-2\,\gravr\,\ssf)}{ [\bha^2 + r\,(r-2\gravr)]
    [r^3 + b^2\,(2\,\gravr\,\ssf^2 - \ssg\,r)]^{1/2} }\ .
\eeq
To understand the sign, consider Figs.~\ref{fig:vangle} and \ref{fig:BHCoords}.  In the case
of retrograde motion, $\kazym_f = -\hat{\vth}$ and $\kazym_i = \pi+\hat{\vth}_S$,
with $\kazym_i > \kazym_f$ (cf. Section~\ref{sec:bend-angle}); recall from Section~\ref{sec:bend-angle} that $\hat{\vth}$ and $\hat{\vth}_S$ are the respective projections onto the $xy$-plane of the angles $\vth$ and $\vth_S$.  For the ``incoming'' ray segment (from
the source the point of closest approach), we have (see, e.g., \cite[p.~189]{weinberg}),
\beqa
  \kazym_i - \kazym_0 = \int_{r_0}^{\infty} \left|\frac{d\kazym}{dr}\right|\,dr\,,\nonumber
\eeqa
where $\kazym_0$ is the value of $\kazym$ at the point of closest
approach.  For the ``outgoing'' segment (from the observer the point of
closest approach), we have
\beqa
  \kazym_0 - \kazym_f = \int_{r_0}^{\infty} \left|\frac{d\kazym}{dr}\right|\,dr\,.\nonumber
\eeqa
Putting them together yields
\beqa \label{eq:app:hint}
  \pi + \hat{\vth}_S + \hat{\vth} = \kazym_i - \kazym_f
  = 2 \int_{r_0}^{\infty} \left|\frac{d\kazym}{dr}\right|\,dr\,.\nonumber
\eeqa
Identifying $\hat{\vth} + \hat{\vth}_S$ as the ``horizontal" bending angle
$\ath$ (for quasi-equatorial lensing; see eqn.~(\ref{bavh})), we can rewrite this equation in the more
familiar form (cf.\ \cite{KP1})
\beq \label{eq:app:hbend}
  \ath = 2 \int_{r_0}^{\infty} \left|\frac{d\kazym}{dr}\right|\,dr
    \ - \ \pi\,.
\eeq
In the case of prograde motion, we have $\kazym_f = \hat{\vth}$ and $\kazym_i = -(\pi+\hat{\vth}_S)$ with $\kazym_f > \kazym_i$.
Similar logic then yields
\beq
  \pi + \hat{\vth}_S + \hat{\vth} = \kazym_f - \kazym_i
  = 2 \int_{r_0}^{\infty} \left|\frac{d\kazym}{dr}\right|\,dr\,.
\eeq
Identifying $\hat{\vth} + \hat{\vth}_S = \ath$ again yields
eqn.~(\ref{eq:app:hbend}).

Thus, eqn.~(\ref{eq:app:hbend}) represents the general expression for
the ``horizontal" component of the bending angle.  The integrand depends on the
invariant impact parameter, $b$, but the integral itself also
depends on the coordinate distance of closest approach, $r_0$.
For pedagogical purposes, and to connect with previous studies
of lensing by Kerr black holes, it is useful to express the
integral purely in terms of $r_0$, and later to convert back
to $b$.

In the weak-deflection regime, $r - 2 \gravr$ and $r - 2\,\gravr\,\ssf$
are always positive, so all factors in eqn.~(\ref{eq:app:hintgnd}) are
positive except for $\sns = \pm1$.  Hence the absolute value in
eqn.~(\ref{eq:app:hbend}) simply removes the factor of $\sns$.  Changing
integration variables to $x = r_0/r$, we can write the bending
angle as
\beqa
  \ath = 2\int_{0}^{1} \frac{ 1 - 2\,\ssf\,h\,x }
    {(1 - 2\,h\,x + \ahat^2\,h^2\,x^2)\ [ \ssg(1-x^2) - 
    2\,\ssf^2\,h(1-x^3) ]^{1/2}}\ 
    dx \ - \ \pi\,,\nonumber
\eeqa
where $h = \gravr/r_0$, and we have used eqn.~(\ref{eq:app:bofr}) to
substitute for $b$ in terms of $r_0$.  In the weak-deflection
regime $h \ll 1$, so we can expand the integrand as a Taylor
series in $h$ and then integrate term by term to obtain
\beqa \label{eq:app:bangle-h-kerr}
  \ath(h) = &\ssc_0&\!\!\pi \ +\ 4\,\ssc_1\,h
    \ + \ \left(-4\,\ssc_2 + \frac{15\pi}{4}\,\ssd_2\right) h^2
    \ + \ \left(\frac{122}{3}\,\ssc_3 - \frac{15\pi}{2}\,\ssd_3\right) h^3\nonumber\\
    &+& \left(-130\,\ssc_4 + \frac{3465\pi}{64}\,\ssd_4\right) h^4 + \order{h}{5} ,\label{eq:app:bangle-h-kerr}
\eeqa
where
\beqan
  \ssc_0 &=& \frac{1}{\ssg^{1/2}} \ - \ 1\,, \\
  \ssc_1 &=& \frac{\ssf^2 + \ssg - \ssf\,\ssg}{\ssg^{3/2}}\ , \\
  \ssc_2 &=& \frac{\ssf^2 (\ssf^2 + \ssg - \ssf\,\ssg)}{\ssg^{5/2}}\ , \\
  \ssd_2 &=& \frac{1}{15\,\ssg^{5/2}}\,\left[ 15\,\ssf^4
    - 4\,\ssg\,(\ssf-1)\,(3\,\ssf^2 + 2\,\ssg)
    - 2\,\ahat^2\,\ssg^2 \right] , \\
  \ssc_3 &=& \frac{1}{61\,\ssg^{7/2}}\,\left[ 61\,\ssf^6
    - \ssg\,(\ssf-1)\,(45\,\ssf^4 + 32\,\ssf^2\,\ssg + 16\,\ssg^2)
    - 4\,\ssg^2\,\ahat^2 (2\,\ssf^2 + 2\,\ssg - \ssf\,\ssg) \right] , \\
  \ssd_3 &=& \frac{\ssf^2}{\ssg}\ \ssd_2\ ,\\
  \ssc_4 &=& \frac{\ssf^2}{65\,\ssg^{9/2}}\,\left[65\,\ssf^6 - 49 (\ssf-1)\,\ssf^4\,\ssg - 8 \ssf^2\,(-4 + \ahat^2 + 4\,\ssf)\,\ssg^2 + 
   4 \left(4 + \ahat^2 (\ssf-2) - 4\,\ssf\right)\,\ssg^3\right]\ ,\\
   \ssd_4 &=& \frac{1}{1155\,\ssg^{9/2}}\left[1155\,\ssf^8 - 840 (\ssf-1)\,\ssf^6\,\ssg - 140\,\ssf^4 (-4 + \ahat^2 + 4\,\ssf)\, \ssg^2 + 
 80 (4 + \ahat^2 (\ssf-2) - 4\,\ssf)\,\ssf^2\,\ssg^3\right.\\ 
 &&~~~~+\left.8 (16 - 12\ahat^2 + \ahat^4 + 8 (\ahat^2-2)\,\ssf)\,\ssg^4\right]\ .
\eeqan
(Terms beyond order four in the bending angle series can be
derived but are not used in our study.)  In the absence of spin,
we have $\ahat = 0$ and $\ssf = \ssg = 1$, so the coefficients become
\beqa
  \ssc_0 = 0, \qquad
  \ssc_1 = \ssc_2 = \ssc_3 = \ssd_2 = \ssd_3 = \ssc_4 = \ssd_4 = 1.\nonumber
\eeqa
In this limit, eqn.~(\ref{eq:app:bangle-h-kerr}) reduces to the correct
Schwarzschild result in \cite{KP1}. 

Let us briefly consider the bending angle to lowest order in
$\gravr/r_0$ and $\bha/r_0$.  At first order, $b \approx r_0$ so
from eqn.~(\ref{eq:FG}) we have $\ssf \approx 1 - \sns\,\bha/r_0$
and $\ssg \approx 1$.  This yields $\ssc_0 \approx 0$ and
$\ssc_1 \approx 1 - \sns\,\bha/r_0$.  So to lowest order
eqn.~(\ref{eq:app:bangle-h-kerr}) gives
\beqa
  \ath \approx 
    4\,\frac{\gravr}{r_0}\,\left(1 - \sns\,\frac{\bha}{r_0} \right) ,\nonumber
\eeqa
which recovers the known result for such a regime
(see., e.g., \cite[p.~281]{boyer-lindquist}).

The expression (\ref{eq:app:bangle-h-kerr}) is coordinate-dependent because it involves the coordinate distance of
closest approach, $r_0$.  We must rewrite the formula in
terms of the impact parameter $b$ to obtain an invariant
result.  We use eqn.~(\ref{eq:rofb}) to write $h = \gravr/r_0$ as
a Taylor series in $\gravr/b$,
\beqa
h  &=& \frac{1}{\ssg^{1/2}} \  \left(\frac{\gravr}{b}\right)
    \ + \ \frac{\ssf^2}{\ssg^2} \  \left(\frac{\gravr}{b}\right)^2
    \ + \ \frac{5 \, \ssf^4}{2 \, \ssg^{7/2}} \  
              \left(\frac{\gravr}{b}\right)^3
    \ + \ \frac{8\, \ssf^6 }{\ssg^5}\  
              \left(\frac{\gravr}{b}\right)^4 \nonumber\\
&&\qquad
    \ + \ \frac{231 \, \ssf^8}{8 \, \ssg^{13/2}}\   
          \left(\frac{\gravr}{b}\right)^5
    \ + \ \frac{112 \, \ssf^{10}}{\ssg^8}\   \left(\frac{\gravr}{b}\right)^6
    \ + \ \order{\frac{\gravr}{b}}{7} ,
    \label{eq:app:hser}
\eeqa
and insert this into
eqn.~(\ref{eq:app:bangle-h-kerr}) to obtain a series expansion for
the bending angle in $\gravr/b$:
\beqa
  \ath(b) &=&  C_0
    \ + \ C_1 \left(\frac{\gravr}{b}\right)
    \ + \ C_2 \left(\frac{\gravr}{b}\right)^2
    \ + \ C_3 \left(\frac{\gravr}{b}\right)^3
    \ + \ C_4 \left(\frac{\gravr}{b}\right)^4
    \ + \ \order{\frac{\gravr}{b}}{5} ,\nonumber\\
    \label{eq:app:bangle-b1-kerr}
\eeqa
where
\beqan
  C_0 &=& \left(\frac{1}{\ssg^{1/2}} - 1 \right)\,\pi\,, \\
  C_1 &=& 4\ \frac{\ssf^2 + \ssg - \ssf\,\ssg}{\ssg^2}\ , \\
  C_2 &=& \frac{\pi}{4\,\ssg^{7/2}}\,\left[ 15\,\ssf^4
    - 4\,\ssg\,(\ssf-1)\,(3\,\ssf^2 + 2\,\ssg)
    - 2\,\ahat^2\,\ssg^2 \right] , \\
  C_3 &=& \frac{8}{3\,\ssg^5}\,\left[ 16 \ssf^6
    - 4\,\ssg\,(\ssf-1)\,(3\,\ssf^4 + 2\,\ssf^2\,\ssg + \ssg^2)
    - \ahat^2\,\ssg^2 (2\,\ssf^2 + 2\,\ssg - \ssf\,\ssg) \right]\ ,\\
  C_4 &=& \frac{3\pi}{64\,\ssg^{13/2}}\,\left[ 1155\,\ssf^8 - 840 (\ssf-1)\ssf^6\,\ssg - 140\,\ssf^4\,(-4 + \ahat^2 + 4\,\ssf)\,\ssg^2\right.\\ 
  &&\left.~~\,+80 \left(4 + \ahat^2 (\ssf-2) - 4\,\ssf\right)\,\ssf^2\,\ssg^3 + 
  8 \left(16 - 12 \ahat^2 + \ahat^4 + 8 (\ahat^2-2)\,\ssf\right)\,\ssg^4 \right] .
\eeqan
Eqn.~(\ref{eq:app:bangle-b1-kerr}) holds for values of $\ssf$
and $\ssg$ where the spin is bounded, $\ahat^2 < 1$.  In other
words, when expanding in $\gravr/b$ we really ought to expand in
$\bha/b$ as well.  Formally, we may accomplish this by writing
$\ssf$ and $\ssg$ in terms of $\ahat$ and $\gravr/b$ as in
eqn.~(\ref{eq:FG}), expanding the coefficients $C_i$ as Taylor series
in $\gravr/b$, and collecting terms to obtain a new series expansion
for the bending angle.  The result is:
\beqa \label{eq:app:bangle-b2-kerr}
  \ath(b) &=& 
          A_1 \left(\frac{\gravr}{b}\right)
    \ + \ A_2 \left(\frac{\gravr}{b}\right)^2
    \ + \ A_3 \left(\frac{\gravr}{b}\right)^3
    \ + \ A_4 \left(\frac{\gravr}{b}\right)^4
    \ + \ \order{\frac{\gravr}{b}}{5} ,\nonumber
\eeqa
where
\beqan \label{eq:app:Ai}
  A_1 &=& 4\ ,\\
  A_2 &=& \frac{15\pi}{4} - 4\,\sns\,\ahat\ ,\\
  A_3 &=& \frac{128}{3} - 10\,\pi\,\sns\,\ahat + 4\,\ahat^2\ ,\\
  A_4 &=& \frac{3465\pi}{64} -192\,\sns\,\ahat + \frac{285\pi\,\ahat^2}{16} - 4\,\sns\,\ahat^3\ .
\eeqan
(Recall that $\ahat^2 < 1$.)  When there is no spin $(\ahat = 0)$, the coefficients reduce to
$A_1 = 4$, $A_2 = 15\pi/4$, $A_3 = 128/3$, and $A_4 = 3465\pi/64$  and recover the
Schwarzschild values in \cite{KP1}.

Note that in our scaled angular variables (\ref{eq:newvar})--(\ref{eq:tseries}), the ``horizontal" bending angle to third order in $\vep$ is
\beqa
\label{eq:horbangle}
\ath(\vep) &=& \frac{4}{\theta_0}\,\vep + \frac{15\pi + 16\,\sns\,\ahat -16\theta_1}{4\theta_0^2}\,\vep^2\nonumber\\
&&~+~\frac{256 + 24\, \ahat^2 - 60\pi\,\sns\,\ahat + 64 D^2 \theta_0^4 - 
 45 \pi \theta_1 + 48\,\sns\,\ahat\,\theta_1 + 24 \theta_1^2 - 
 24 \theta_0 \theta_2}{6\theta_0^3}\,\vep^3 + \order{\vep}{4}\ .\nonumber\\
 &&\label{eq:horbangle}
\eeqa

\subsection{``Vertical" Bending Angle}
\label{sec:app:vertical}

This section presents new results on the vertical component of the bending angle in quasi-equatorial lensing.  From eqns.~(\ref{eq:app:rdot2}) and (\ref{eq:app:edot2}), the 
quasi-equatorial light
bending in the $\kpolar$-direction is governed by the equation
of motion
\beqa
  \frac{d\kpolar}{dr} = \frac{\hat{\dot{\kpolar}}}{\hat{\dot{r}}}
  = \pm i(r)\,\left[\frac{(\dvphi)^2}{\ssg} - \kpolar^2\right]^{1/2} ,\nonumber
\eeqa
where
\beq \label{eq:app:little-i}
  i(r) = \frac{b\,\ssg^{1/2}}
    {(r^4 - b^2\,\ssg\,r^2 + 2\gravr\,b^2\,\ssf^2\,r)^{1/2}}\ .
\eeq
The equation of motion has solutions of the form
\beqa
  \kpolar(r) = \frac{\dvphi}{\ssg^{1/2}}\ \sin\left[ \pm I(r) + p \right]\,,\nonumber
\eeqa
where
\beq \label{eq:app:big-I}
  I(r) = \int_{r_0}^{r} i(r')\,dr'\,,
\eeq
and $p$ is a constant of integration.  We are interested in the
two asymptotic values ($r \to \infty$),
\beqa
  \kpolar_\pm = \frac{\dvphi}{\ssg^{1/2}}\ \sin\left( \pm I_\infty + p \right)\,,\nonumber
\eeqa
where $I_\infty = \lim_{r\to\infty} I(r)$.  We can eliminate $p$
and relate the two solutions to one another:
\beq \label{eq:app:etaminus}
  \kpolar_- = - \frac{\dvphi}{\ssg^{1/2}}\ \sin\left[ -2 I_\infty
    + \sin^{-1}\left( \frac{\ssg^{1/2}}{\dvphi}\ \kpolar_+ \right) \right] .
\eeq
The asymptotic values $\kpolar_\pm$ must correspond to the initial
and final values, $\kpolar_i$ and $\kpolar_f$, introduced in
Appendix~\ref{app:sec:BHCoords}, but we must determine the correspondence.
In order to do that, we first examine $\kpolar_i$ and
$\kpolar_f$ more carefully, using eqn.~(\ref{eq:app:tp2ex2}).  Recall that
in the quasi-equatorial regime we have $\vphi = \vphi_0 + \dvphi$ and 
$\vphi_S = \vphi_0 + \pi + \dvphi_S$, with $\vphi_0 = \pi$ for prograde motion and $\vphi_0 = 0$ for retrograde motion.  Using these relations, eqns.~(\ref{eq:app:ToEtaXi}) and (\ref{eq:app:FromEtaXi}) both become
\beqa
  \sin\kpolar_i = \sin\vth_S\,\sin\dvphi_S, \quad
  \sin\kpolar_f = \sin\vth\,\sin\dvphi  \,.\nonumber
\eeqa
Since we are working to first order in $\kpolar$ and $\dvphi$,
we can write these as
\beqa
  \kpolar_i = \dvphi_S\ \sin\vth_S\,, \qquad
  \kpolar_f = \dvphi  \ \sin\vth\,.\nonumber
\eeqa
Upon considering the spherical case (see below), we recognize
that we want to put $\kpolar_- = \kpolar_i$ and $\kpolar_+ = \kpolar_f$
in eqn.~(\ref{eq:app:etaminus}).  This substitution yields
\beq \label{eq:app:W1}
  \dvphi_S = \frac{\dvphi}{\ssg^{1/2} \sin\vth_S}\ 
    \sin\left[ -2 I_\infty + \sin^{-1}\left( \ssg^{1/2} \sin\vth \right)
    \right] 
  \equiv \frac{W(\vth)}{\sin\vth_S}\dvphi\,.
\eeq
Notice that the coefficient of $\dvphi$ depends only on $\vth$, not on
$\vphi$.  (In addition to the explicit $\vth$ dependence, there
is implicit dependence through $\vth_S$ and 
$I_\infty$, which depends on $b = d_L \sin\vth$.)  We can therefore
define it to be the function $W(\vth)$, with a factor of
$\sin\vth_S$ that will prove to be convenient later.

Before evaluating $I_\infty$, let us check the case of a spherical
lens to make sure our result is reasonable.  For a spherical lens,
$\bha = 0$ and $\ssf = \ssg = 1$, so we have
\beqa
  2 I_\infty = 2 \int_{r_0}^{\infty} \frac{b\,dr}
    {r^{1/2} [r^3 - b^2 (r-2\gravr)]^{1/2}}
  = \pi + \vth_S + \vth\,,\nonumber
\eeqa
where the last equality is obtained after comparison with the
spherical limits of eqns.~(\ref{eq:app:hintgnd}) and (\ref{eq:app:hint}).
Together with our choices $\kpolar_- = \kpolar_i$ and $\kpolar_+ = \kpolar_f$, eqn.~(\ref{eq:app:W1}) then becomes
\beqa
\label{eq:app:W1-sph}
  \dvphi_S = \frac{\dvphi}{\sin\vth_S}\ 
    \sin\left[ - \pi - \vth_S - \vth + \sin^{-1}(\sin\vth) \right]
  = \dvphi\,,\nonumber
\eeqa
which is consistent with the symmetry.
This
verifies our choice of signs above.

We now evaluate the integral (for the general case, not just
the spherical limit), in parallel with the analysis in
Appendix~\ref{sec:app:horizontal}.  From eqns.~(\ref{eq:app:little-i}) and (\ref{eq:app:big-I}) we have
\beqa
  I_\infty = \int_{r_0}^{\infty} \frac{b\,\ssg^{1/2}}
    {(r^4 - b^2\,\ssg\,r^2 + 2\gravr\,b^2\,\ssf^2\,r)^{1/2}}\ dr\,.\nonumber
\eeqa
Using eqn.~(\ref{eq:app:bofr}) for $b$ and changing integration variables to $x = r_0/r$ yields
\beqa
  I_\infty = \int_{0}^{1} \frac{\ssg^{1/2}}
    {[\ssg(1-x^2) - 2\,\ssf^2\,h (1-x^3)]^{1/2}}\ dx\,,\nonumber
\eeqa
where $h = \gravr/r_0$.  Taylor expanding in $h$ and
integrating term by term gives
\beqa
  I_\infty = \frac{\pi}{2}
    + \frac{2\,\ssf^2}{\ssg}\,h
    + \frac{\ssf^4}{8\,\ssg^2}(15\pi-16)\,h^2
    + \frac{\ssf^6}{12\,\ssg^3}(244-45\pi)\,h^3
    + \frac{5\ssf^8}{128\,\ssg^4}(-1664+693\pi)\,h^4 +\order{h}{5} .\nonumber
\eeqa
We now use eqn.~(\ref{eq:app:hser}) to write $h$ in terms of $\gravr/b$,
and then collect terms to obtain
\beqa
  I_\infty = \frac{\pi}{2}
    \ + \ \frac{2\,\ssf^2}{\ssg^{3/2}} \left(\frac{\gravr}{b}\right)
    \ + \ \frac{15\pi\,\ssf^4}{8\,\ssg^3} \left(\frac{\gravr}{b}\right)^2
    \ + \ \frac{64\,\ssf^6}{3\,\ssg^{9/2}} \left(\frac{\gravr}{b}\right)^3
    \ + \ \frac{3465\pi\,\ssf^8}{128\,\ssg^{6}} \left(\frac{\gravr}{b}\right)^4
    \ + \ \order{\frac{\gravr}{b}}{5} .\nonumber
\eeqa
As in eqn.~(\ref{eq:app:bangle-b2-kerr}), when we expand in $\gravr/b$
we ought to expand in $\bha/b$ as well.  We use eqn.~(\ref{eq:FG})
to write $\ssf$ and $\ssg$ in terms of $\ahat$ and $\gravr/b$,
and then collect terms to find
\beqa
\label{eq:app:Iseries}
  I_\infty &=& \frac{\pi}{2}
    \ + \ 2 \left(\frac{\gravr}{b}\right)
    \ + \ \left(\frac{15\pi}{8} - 4\,\sns\,\ahat\right)
          \left(\frac{\gravr}{b}\right)^2
    \ + \ \left(\frac{64}{3} - \frac{15\pi\,\sns\,\ahat}{2} + 5\ahat^2 \right)
          \left(\frac{\gravr}{b}\right)^3\nonumber\\
          &&~+\left(\frac{3465\pi}{128} - 128\,\sns\,\ahat + \frac{135\pi\,\ahat^2}{8} -6\,\sns\,\ahat^3\right)\left(\frac{\gravr}{b}\right)^4\ + \ \order{\frac{\gravr}{b}}{5} .
\eeqa
This is to be used with eqn.~(\ref{eq:app:W1}) to describe the ``vertical"
bending (see eqn.~(\ref{eq:lens4v}) in Section~\ref{subsec:quasi} above).  Note also that the expression inside the square root in eqn.~(\ref{squareroot}) is
\beqa
\label{squareroot2}
1- W(\vth)^2 &=& 1- \left(\frac{1}{\ssg^{1/2}}\,\sin\left[ -2 I_\infty + \sin^{-1}\left( \ssg^{1/2} \sin\vth \right)
    \right]\right)^2 \nonumber\\
    &=& \left(1 - \sin^2\vth\right) + 4\sin2\vth\,\left(\frac{\gravr}{b}\right) - \left[16\cos2\vth + \left(\!-\frac{15\pi}{4} + 8\,\sns\,\ahat\right)\sin2\vth\right]\,\left(\frac{\gravr}{b}\right)^2\nonumber\\
    &&~-\biggl[(30 \pi - 64\,\sns\,\ahat) \cos2\vth + 
  \ahat\,(15\pi\,\sns-10\,\ahat) \sin2\vth - 
     4\,\ahat^2\tan\vth\biggr]\,\left(\frac{\gravr}{b}\right)^3 + \order{\frac{\gravr}{b}}{4}.
\eeqa
Since $0 < \vth < \pi/2$ and $\gravr/b \ll 1$, eqn.~(\ref{squareroot2}) is nonnegative.

Finally, analogously to the ``horizontal" component of the bending angle derived in Section~\ref{sec:bend-angle}, we derive the ``vertical" component of the bending angle, as follows.  Consider the angles $\nu_i$ and $\nu_f$ shown in \reffig{vangle}.  We define $\nu_f$ to be strictly nonnegative and within the interval $[0,\pi/2)$, but allow $\nu_i$ to be negative, so that $-\pi/2 < \nu_i < \pi/2$, and enforce the following sign convention for $\nu_i$.  As shown in \reffig{vangle}, $\nu_i$ is the angle whose vertex is the point $B'$ on the lens plane, and is measured from a line parallel to the equatorial plane.  If $\nu_i$ goes {\it away} from the equatorial plane, then we take it to be positive; otherwise it is negative (e.g., the $\nu_i$ shown in \reffig{vangle} is positive).  Now denote by $\hat{\nu}_i$ and $\hat{\nu}_f$ the respective projections onto the $xz$-plane of the angles $\nu_i$ and $\nu_f$, and adopt the same sign conventions for them.  With these conventions, the ``vertical" component of the bending angle can be unambiguously expressed as 
$$
\atv = \hat{\nu}_f - \hat{\nu}_i\ .
$$
By the positivity of $\hat{\nu}_f$ and the fact that the bending is nonnegative, we have
$$
\hat{\nu}_i \leq \hat{\nu}_f\ .
$$
(Indeed, with our signs conventions the condition $\hat{\nu}_i > \hat{\nu}_f$ would be equivalent to repulsion of the light ray.)  Writing $\hat{\nu}_i$ and $\hat{\nu}_f$ in terms of the angles $\vth, \vphi, \vth_S, \vphi_S$, we have
\beqa
\hat{\nu}_f &=& \tan^{-1}(\tan\vth\sin\vphi)\ ,\nonumber\\
\hat{\nu}_i &=& \tan^{-1}(\tan\vth_S\sin(\pi-\vphi_S))\ ,\nonumber
\eeqa
which in the quasi-equatorial regime reduce to
\beqa
\hat{\nu}_f &\approx& \pm\dvphi\tan\vth\ ,\nonumber\\
\hat{\nu}_i &\approx& \mp\dvphi_S\tan\vth_S\nonumber\ ,
\eeqa
where we have set $\vphi = \vphi_0 + \dvphi$, $\dvphi_S = \vphi_0 + \pi + \dvphi_S$, with $\vphi_0 = 0$ (retrograde motion) or $\pi$ (prograde motion), and expanded to linear order in the small angles $\dvphi$ and $\dvphi_S$.  Using the identities $W(\vth)\,\dvphi = \sin\vth_S\,\dvphi_S$ and $\vth_S = \ath - \vth$ given by eqns.~(\ref{eq:app:W1}) and (\ref{bavh}), we can thus write $\atv$ as
\beqa
\atv \approx \pm\,\dvphi\left[\tan\vth + \frac{W(\vth)}{\cos(\ath - \vth)}\right]\ .\nonumber
\eeqa
The expression inside the square brackets is of the form $16D\,\csc\vth\,\sec^2\vth\,\vep^2 + \order{\vep}{4}$, so it is positive (recall that $0 < \vth < \pi/2$).  Since the bending angle is strictly nonnegative, we will adopt $``+"$ for $\dvphi \geq 0$ and $``-"$ for $\dvphi < 0$, so that we may write 
\beqa
\label{vert1}
\atv \approx \dvphi\left[\tan\vth + \frac{W(\vth)}{\cos(\ath - \vth)}\right]\ .
\eeqa
We now expand eqn.~(\ref{vert1}) in our scaled angular variables (\ref{eq:newvar})--(\ref{eq:tseries}) to third order in $\vep$ to obtain
\beqa
\atv(\vep) &\approx& \dvphi\,\Biggl\{\tan\vth + \frac{1}{\ssg^{1/2}}\sin\left[ -2 I_\infty + \sin^{-1}\left( \ssg^{1/2} \sin\vth \right)
    \right]\frac{1}{\cos(\ath-\vth)}\Biggr\}\nonumber\\
&=& \dvphi\,\Biggl\{\frac{4}{\theta_0}\,\vep + \frac{15\pi - 32\,\sns\,\ahat -16\theta_1}{4\theta_0^2}\,\vep^2\nonumber\\
&&~+\frac{256 + 72\,\ahat^2 - 90\pi\,\sns\,\ahat + 64 D^2 \theta_0^4 + 
 96\,\sns\,\ahat\,\theta_1 - 45\pi \theta_1 + 24 \theta_1^2 - 
 24 \theta_0 \theta_2}{6\theta_0^3}\,\vep^3\nonumber\\ &&~+\order{\vep}{4}\Biggr\}\ .\label{eq:bavy}
\eeqa
The result in eqn.~(\ref{eq:bavy}) is new.



\section{Quasi-Equatorial Time Delay}
\label{app:Kerr-Tdel}

We now compute the time delays for quasi-equatorial lensed images.
Let $R_{\rm src}$ and $R_{\rm obs}$ be the radial coordinates of
the source and observer, respectively.  From geometry relative
to the flat metric of the distant observer, who is assumed to
be at rest in the Boyer-Lindquist coordinates, we can work out 
\beqa \label{eq:app:Rdef}
  R_{\rm obs} = d_L\,, \qquad
  R_{\rm src} = \left( d_{LS}^2 + d_S^2 \tan^2 \cb \right)^{1/2} .\nonumber
\eeqa
The radial distances are very nearly the same as angular
diameter distances since the source and observer are in the
asymptotically flat region of the spacetime.  In the absence
of the lens, the spacetime would be flat and the light ray
would travel along a linear path of length $d_S/\cos\cb$
from the source to the observer.

The time delay is the difference between the light travel
time for the actual ray, and the travel time for the ray the
light would have taken had the lens been absent.  This can be
written as
\beqa \label{eq:app:tdel}
  c \tau = T(R_{\rm src}) + T(R_{\rm obs}) - \frac{d_S}{\cos\cb}\ ,\nonumber
\eeqa
with
\beqa
  T(R) = \int_{r_0}^{R} \left|\frac{dt}{dr}\right|\,dr
  = \int_{r_0}^{R} \left|\frac{\dot{t}}{\dot{r}}\right|\,dr\,.\nonumber
\eeqa
We use $\dot{t}$ and $\dot{r}$ from eqns.~(\ref{eq:app:tdot2}) and
(\ref{eq:app:rdot2}), substitute for $b$ using eqn.~(\ref{eq:app:bofr}), and
change integration variables to $x = r_0/r$.  This yields
\beqa
  T(R) = r_0 \int_{r_0/R}^{1} \frac
    { (\ssg - 2\,\ssf^2\,h)^{1/2} (1 + \ahat^2\,h^2\,x^2)
      - 2\,\sns\,\ahat\,\ssf\,h^2\,x^3 }
    { x^2\,(1 - 2\,h\,x + \ahat^2\,h^2\,x^2)\ 
      [ \ssg(1-x^2) - 2\,\ssf^2\,h(1-x^3) ]^{1/2} }\ dx\,,\nonumber
\eeqa
where $h = \gravr/r_0$.  We expand the integrand as a Taylor
series in $h$ and integrate term by term.  The result is a
series in $h$ whose coefficients are rational functions of
$\omega = r_0/R$.  The first three terms in the expansion are
\beqa
  T(R) &=& \sqrt{R^2-r_0^2}
    \ + \ h\,r_0\,\left[ \frac{\ssf^2 \sqrt{1-\omega^2}}{\ssg (1+\omega)}
          + 2 \ln\left(\frac{1+\sqrt{1-\omega^2}}{\omega}\right) \right]
\label{eq:app:Tser} \\
  && + \ h^2\,r_0\,\Biggl[
    \frac{3\,\ssf^4 + 4\,\ssf^2\,\ssg + 8\,\ssg^2}{2\,\ssg^2}
    \left( \frac{\pi}{2} - \sin^{-1}\omega \right)
      - 2\,\sns\,\ahat\,\ssf\,\ssg^{-1/2}\,\sqrt{1-\omega^2} \nonumber\\
  &&\qquad\qquad
      -\, \frac{\ssf^2\,(4\,\ssg+(\ssf^2+4\,\ssg)\omega)\,\sqrt{1-\omega^2}}{2\,\ssg^2\,(1+\omega)^2} 
    \Biggr]
  \ + \ \order{h}{3}\ .\nonumber
\eeqa
The third-order term is easily obtained, and is needed in the derivation of eqns.~(\ref{T2}) and (\ref{D2}), but is too unweildy to write here.  Note that if
we substitute for $\ssf$ and $\ssg$ using eqn.~(\ref{eq:FG}) and take the
far-field limit, we recover previous results (e.g., \cite{dymnikova,dymnikova2}).

\section{Third-Order Results}
\label{app:third-order}
\noindent The vanishing of the third-order term in the ``horizontal" lens equation (\ref{eq:lens3h}) yields
\beqa
\label{eq:theta3}
\theta_3 &=& \frac{1}{768\theta_0^2\,(\theta_0^2 + 1)}\Biggl\{128\,\sns\,\ahat\left[-384 + 9 (5 \pi - 2 \theta_1) \theta_1 + 
   4 \theta_0 \left(3 \theta_2 + 
      8 \theta_0 (2 D (3 - 2 D \theta_0^2)\right.\right.\nonumber\\
      &&\qquad~+\left.\left. 
         3 (-1 + D) (-1 + D \theta_0^2)\,\xi)\right)\right] + 768\,\ahat^2\,(2 \pi - 3 \theta_1) -768\,\sns\,\ahat^3\nonumber\\
         &&\qquad~+15 \pi \left[3 (487 + 48 \theta_1^2) + 
   32 \theta_0 \left(-3 \theta_2 + 
      8 \theta_0 (2 D (-3 + 2 D \theta_0^2) - 
         3 (-1 + D) (-1 + D \theta_0^2) \xi)\right)\right]\nonumber\\
         &&\qquad~+ 256 \theta_1 \left[-3 (48 + \theta_1^2) + 
   2 \theta_0 \left(3 \theta_2 + 
      4 \theta_0 (D (6 + D \theta_0^2 (7 - 6 \theta_0^2)) - 
         3 (-1 + D) (1 + 2 D \theta_0^2) \xi)\right)\right]\Biggr\}\ .\nonumber
\eeqa
The third-order magnification term in eqn.~(\ref{eq:mag}) is
\beqa
\mu_3 = \frac{1}{12288\,(\theta_0^2 - 1)^4 (\theta_0+1)^7}\Bigg[\ahat\,{\tt M_1} + \ahat^2\,{\tt M_2} + \ahat^3\,{\tt M_3} + {\tt M_4}\Bigg]\ ,\nonumber
\eeqa
where
\beqa
{\tt M_1} &=& 32\,\sns\,\theta_0 \biggl[-675 \pi^2 (\theta_0^2-1)^2 (-1 - 6 \theta_0^2 - 
    14 \theta_0^4 + 10 \theta_0^6 - 17 \theta_0^8 + 
    24 \theta_0^{10})\nonumber\\
    &&~+ 
 2048 (\theta_0^4-1)^2 (-3 + 
    3 (-1 + D) D \theta_0^{12} \xi + \theta_0^{10} \left(2 (-3 + D) D\right.\nonumber\\
    &&~+\left. 
       3 (-1 + D) (1 + 4 D)\,\xi\right) + \theta_0^2 (-3 (4 + \xi) + 
       D (-6 + 2 D + 3 \xi))\nonumber\\
       &&~+ 
    2 \theta_0^6 (-6 - 9 \xi - 9 D (2 + \xi) + 
       2 D^2 (8 + 9 \xi)) + \theta_0^8 (33 + 6 \xi - 
       6 D (-2 + D (4 + \xi)))\nonumber\\
       &&~+ 
    3 \theta_0^4 (2 (3 + \xi) + D (4 + 5 \xi - D (8 + 7 \xi)))\biggr]\ ,\nonumber\\
{\tt M_2} &=& 384\pi\,\theta_0\left[(\theta_0^2-1) (15 + 95 \theta_0^2 + 231 \theta_0^4 - 
   1797 \theta_0^6 + 1777 \theta_0^8 - 1287 \theta_0^{10} + 
   2217 \theta_0^{12} + 29 \theta_0^{14})\right]\ ,\nonumber\\
{\tt M_3} &=& -49152\,\sns\,\theta_0^7\left(5-6\theta_0^2 + 18\theta_0^4 -6\theta_0^6 + 5\theta_0^8\right)\ ,\nonumber\\
{\tt M_4} &=& 45 \pi \theta_0 (\theta_0^2-1) \left\{-16 (-1 + \theta_0^4)^2 \left[-25
- 128 D^2 + (-125 + 128 (4 - 7 D) D) \theta_0^2\right.\right.\nonumber\\
&&~+\left.\left. (-1711 + 
         128 D (-8 + 5 D)) \theta_0^4 + (1461 + 
         128 (8 - 15 D) D) \theta_0^6 + 
      256 (-2 + D) D \theta_0^8\right]\right.\nonumber\\
      &&~-\left. 
   225 \pi^2 (-1 + \theta_0^2)^2 (1 + 
      7 (\theta_0^2 + 3 \theta_0^4 + 5 \theta_0^6))\right.\nonumber\\
      &&~-\left. 
   4096 (-1 + 
      D) (\theta_0 - \theta_0^5)^2 (-1 + \theta_0^2 (2 - 
         2 \theta_0^2 + \theta_0^4 + 
         D (9 + \theta_0^2 + 5 \theta_0^4 + \theta_0^6))) \xi\right\}\ .\nonumber
\eeqa
(We will forgo writing $\mu_3^{\pm}$ as the expressions are too lengthy.)  Expressed in terms of $\beta$, the third-order relations are
\beqa
\mu_3^+\ +\ \mu_3^- &=& -\frac{\sns^+\,\ahat\,}{128}\frac{225 \pi^2 - 2048 (1 + (-1 + D) D\,\xi)}{|\beta|}\nonumber\\
      &&- \ \frac{\pi\,\ahat^2}{32}\frac{(-7168 - 1224 \beta^2 + 1272 \beta^4 + 
   230 \beta^6 + 15 \beta^8)}{\beta^2\,(\beta^2+4)^{7/2}} + \frac{15\pi\,\tt M_5}{4096\,(\beta^2+4)^{7/2}}\ , \nonumber\\
    \mu_3^+\ -\ \mu_3^- &=&\frac{\sns^+\,\ahat}{384}\frac{{\tt M_6}}{\beta^2\,(\beta^2+4)^{7/2}}\ +\ \frac{5\pi\,\ahat^2}{32}\frac{(3\beta^2 +4)}{|\beta|^3}\ -\ 8\,\sns^+\,\ahat^3\,\frac{16+14+5\beta^4}{\beta^4\,(\beta^2+4)^{7/2}}\nonumber\\
    &&- \ \frac{15\pi\,(25 (-16 + 9 \pi^2) + 
   2048 D (-D + 2 (-1 + D) \xi))}{4096\,|\beta|}\ ,\label{eq:mu3}
\eeqa
where
\beqa
{\tt M_5} &=&225 \pi^2 (140 + 70 \beta^2 + 14 \beta^4 + \beta^6) - 
 16 (4 + \beta^2) \left(3672 + 250 \beta^2 + 25 \beta^4 - 
    512 (1 + \beta^2)\,\xi\right)\nonumber\\
    &&~+ 
 4096 D (4 + \beta^2) \left(4 + 
    10 \xi + \beta^2 (4 + (8 + \beta^2) \xi)\right)\nonumber\\
    &&~- 
 2048 D^2 (4 + \beta^2) \left(12 + 
    24 \xi + \beta^2 (14 + \beta^2 + 2 (10 + \beta^2)\,\xi)\right)\ ,\nonumber\\
{\tt M_6} &=&-6144 D (4 + \beta^2) \left(8 (2 + \xi) + \beta^2 (2 + \beta^2) \
(4 + (6 + \beta^2)\,\xi)\right)\nonumber\\
&&~+ 
   2048 D^2 (4 + \beta^2) \left(-24 + 
      48 \xi + \beta^2 (4 (-8 + \beta^2) + 
         3 (2 + \beta^2) (8 + \beta^2)\,\xi)\right)\nonumber\\
         &&~+ 
   3 \left[225 \pi^2 (8 + 92 \beta^2 + 70 \beta^4 + 
         14 \beta^6 + \beta^8)\right.\nonumber\\
         &&~-\left. 
      2048 (4 + \beta^2) (\beta^6 + 8 (-2 + \xi) + 
         4 \beta^2 (2 + \xi) + 2 \beta^4 (5 + \xi))\right]\ .\nonumber
\eeqa

\noindent Finally, the third-order centroid term is
\beqa
  \Theta_{{\rm cent},3} = \frac{\ahat\,{\tt C_2} + \ahat^2\,{\tt C_3} + \ahat^3\,{\tt C_4} + {\tt C_5}}{12288\,\beta^2\,(\beta^2+2)^4(\beta^2+4)^{5/2}}\ ,\nonumber
    \eeqa
where
   \beqa
           {\tt C_2} &=&16 s \beta^2 (2 + \beta^2) (480 \pi (2 + \beta^2)^2 (4 + \
\beta^2) (14 + 5 \beta^2) + 
   \nonumber\\
   &&~+ 225 \pi^2\sqrt{
    4 + \beta^2} (1168 + 2016 \beta^2 + 1642 \beta^4 + 
      597 \beta^6 + 97 \beta^8 + 6 \beta^{10})\nonumber\\
      &&~- 
   2048 (4 + \beta^2)^{
    3/2} (6 (\beta^8 + 
         16 (4 + \xi) + \beta^6 (12 + \xi) + \beta^4 (46 + 
            5 \xi) + \beta^2 (68 + 6 \xi))\nonumber\\
            &&~- 
      6 D (-32 + \beta^6 (-2 + \xi) + 48 \xi + 
         5 \beta^4 (-2 + 3 \xi) + 2 \beta^2 (-6 + 29 \xi))\nonumber\\
         &&~+ 
      D^2 (-5 \beta^6 + 60 \beta^4 \xi + 32 (-5 + 6 \xi) + 
         4 \beta^2 (17 + 78 \xi))))\ ,\nonumber\\
         {\tt C_3} &=& 384 \beta (2 + \beta^2) (-32 \beta^2 (2 + \beta^2)^2 (4 + \
\beta^2)^{
    3/2}\nonumber\\
    &&~+ \pi (-18176 - 15024 \beta^2 + 1512 \beta^4 + 
      5172 \beta^6 + 1926 \beta^8 + 280 \beta^{10} + 
      15 \beta^{12}))\ ,\nonumber\\
      {\tt C_4} &=& -24576 s \beta^2 \sqrt{
 4 + \beta^2} (-32 + 84 \beta^2 + 44 \beta^4 + 5 \beta^6)\ ,\nonumber\\
 {\tt C_5} &=& -\beta^3 (2 + \beta^2)^2 (10800 \pi^2 \sqrt{
    4 + \beta^2} (8 + 6 \beta^2 + \beta^4) + 10125 \pi^3 (268 + 280 \beta^2 + 105 \beta^4 + 
      17 \beta^6 + \beta^8)\nonumber\\
      &&~- 
   32768 (2 + \beta^2) (4 + \beta^2)^{
    5/2} (-D (-6 + D (8 + \beta^2)) + 
      3 (-1 + D) (-1 + 2 D (2 + \beta^2))\,\xi)\nonumber\\
      &&~+ 
   240 \pi (4 + \beta^2) (-3 (4472 + 3198 \beta^2 + 
         556 \beta^4 + 25 \beta^6) + 
      768 D (12 + 2 \beta^4 + \beta^2 (10 - 3 \xi) - 
         10 \xi)\nonumber\\
         &&~+ 768 (6 + 5 \beta^2 + \beta^4) \xi - 
      128 D^2 (188 + 3 \beta^6 - 24 \xi + 
         6 \beta^4 (7 + \xi) + 4 \beta^2 (40 + 3 \xi))))\ .\nonumber
\eeqa


{}


\begin{thebibliography}{}

\bibitem{AKP}
Aazami, A. B., Keeton, C. R., and Petters, A. O., 
``Lensing by Kerr black holes I.  General lens equation and magnification formula,''
J. Math. Phys. (2011).

\bibitem{boyer-lindquist}
Boyer, R. and Lindquist, R.,
``Maximal analytic extension of the Kerr metric,''
J. Math. Phys. {\bf 8}, 265 (1967).

\bibitem{Bozza}
Bozza, V.,
``Quasiequatorial gravitational lensing by spinning black holes in the strong field limit,''
Phys. Rev. D {\bf 67}, 103006 (2003).

\bibitem{BozzaN}
Bozza, V.,
``Extreme gravitational lensing by supermassive black holes,''
Nuovo Cim. {\bf 122B}, 547 (2007).

\bibitem{Bozza2}
Bozza, V.,
``Comparison of approximate gravitational lens equations and a
proposal for an improved new one,''
Phys. Rev. D {\bf 78}, 103005 (2008).

\bibitem{BS2}
Bozza, V. and Scarpetta, G.,
``Strong deflection limit of black hole gravitational
lensing with arbitrary source distances,''
Phys. Rev. D {\bf 76}, 083008 (2007).

\bibitem{BS}
Bozza, V. and Sereno, M.,
``Weakly perturbed Schwarzschild lens in the strong
deflection limit,''
Phys. Rev. D {\bf 73}, 103004 (2006).

\bibitem{Betal2}
Bozza, V., De Luca, F., and Scarpetta, G.,
``Kerr black hole lensing for generic
observers in the strong deflection limit,''
Phys. Rev. D {\bf 74}, 063001 (2006).

\bibitem{Betal}
Bozza, V., De Luca, F., Scarpetta, G., and Sereno, M.,
``Analytic Kerr black
hole lensing for equatorial observers in the strong deflection limit,''
Phys. Rev. D {\bf 72}, 083003 (2005).

\bibitem{chan}
Chandrasekhar, S., 
{\em The Mathematical Theory of Black Holes}
(Clarendon, Oxford, 1983).

\bibitem{dymnikova}
Dymnikova, I.,
``The effect of the relative delay of rays focused by a rotating
massive body,''
Soviet Phys.-JETP Lett. {\bf 59}, 223 (1984).

\bibitem{dymnikova2}
Dymnikova, I.,
``Motion of particles and photons in the gravitational field of
a rotating body,''
Soviet Phys.-Uspekhi {\bf 29}, 215 (1986).

\bibitem{frittelli2}
Frittelli, S., Kling, T. P., and Newman, E. T.,
``Spacetime perspective of Schwarzschild lensing,'' 
Phys. Rev. D {\bf 61}, 064021 (2000).

\bibitem{frittelli}
Frittelli, S. and Newman, E. T.,
``Exact universal gravitational lensing equation,''
Phys. Rev. D {\bf 59}, 124001 (1999).

\bibitem{Iyer1}
Iyer, S. V. and Hansen, E. C.,
``Strong and Weak Deflection of Light in the Equatorial
Plane of a Kerr Black Hole,'' 
gr-qc/0908.0085 (2009).

\bibitem{Iyer2}
Iyer, S. V. and Hansen, E. C.,
``Light's bending angle in the equatorial plane of a
Kerr black hole,'' 
Phys. Rev. D {\bf 80}, 124023 (2009).

\bibitem{KP1}
Keeton, C. R. and Petters, A. O.,
``Formalism for testing theories of gravity using
lensing by compact objects: Static, spherically symmetric case,''
Phys. Rev. D {\bf 72}, 104006 (2005).

\bibitem{KP2}
Keeton, C. R. and Petters, A. O.,
``Formalism for testing theories of gravity using
lensing by compact objects. II. Probing post-post-Newtonian metrics,''
Phys. Rev. D {\bf 73}, 044024 (2006).

\bibitem{KP3}
Keeton, C. R. and Petters, A. O.,
``Formalism for testing theories of gravity using
lensing by compact objects. III. Braneworld gravity,''
Phys. Rev. D {\bf 73}, 104032 (2006).

\bibitem{RB}
Rauch, K. and Blandford, R.,
``Optical caustics in a kerr spacetime and
the origin of rapid x-ray variability in active galactic nuclei,''
Astro. Phys. J. {\bf 421}, 46 (1994).

\bibitem{SerenoDeLuca}
Sereno, M. and De Luca, F.,
``Analytical Kerr black hole lensing in the weak
deflection limit,''
Phys. Rec. D {\bf 74}, 123009 (2006).

\bibitem{SerenoDeLuca2}
Sereno, M. and De Luca, F.,
``Primary caustics and critical points behind a
Kerr black hole,''
Phys. Rec. D {\bf 78}, 023008 (2008).

\bibitem{VqzE}
V\'asquez, S. and Esteban, E.,
``Strong field gravitational lensing by a Kerr
black hole,''
Nuovo Cim. {\bf 119B}, 489 (2004).q

\bibitem{wald}
Wald, R. M., 
{\em General Relativity} (University of Chicago Press, Chicago, 1984).

\bibitem{weinberg}
Weinberg, S., 
{\em Gravitation and Cosmology} (Wiley, New York, 1972).

\bibitem{WernerPetters}
Werner, M. C. and Petters, A. O.,
``Magnification relations for Kerr lensing and
testing cosmic censorship,''
Phys. Rev. D   76, 064024 (2007).



\end{thebibliography}
\end{document}